\documentclass[fleqn,usenatbib]{mnras}

\usepackage{newtxtext,newtxmath}
\usepackage{natbib}
\usepackage[T1]{fontenc}
\newcommand{\orcid}[1]{\href{https://orcid.org/#1}{\textcolor[HTML]{A6CE39}{\aiOrcid}}}

\DeclareRobustCommand{\VAN}[3]{#2}
\let\VANthebibliography\thebibliography
\def\thebibliography{\DeclareRobustCommand{\VAN}[3]{##3}\VANthebibliography}

\usepackage{graphicx}	
\usepackage{amsmath}	

\title[Extraction of cosmic web filaments with WEAVE]{Forecasting the success of the WEAVE Wide-Field Cluster Survey on the extraction of the cosmic web filaments around galaxy clusters}
\author[D. Cornwell et al.]{\parbox{\textwidth}{
  Daniel J. Cornwell$^{1}$\thanks{E-mail: daniel.cornwell@nottingham.ac.uk}, 
  Ulrike Kuchner$^{1}$,
  Alfonso Arag\'{o}n-Salamanca$^{1}$, 
  Meghan E. Gray$^{1}$, 
  Frazer R. Pearce$^{1}$, 
  J. Alfonso L. Aguerri$^{2,3}$,
  Weiguang Cui$^{4,5}$,
J. Méndez-Abreu$^{2,3}$,
Luis Peralta de Arriba$^{6,7}$,
  Scott C. Trager$^{8}$
  } 
 \vspace{0.4cm}
\\
\parbox{\textwidth}{
$^{1}$School of Physics and Astronomy, University of Nottingham, Nottingham NG7 2RD, UK\\
$^{2}$Instituto de Astrofísica de Canarias, Calle Vía Láctea s/n, E-38205 La Laguna, Tenerife, Spain\\
$^{3}$ Departamento de Astrofísica, Universidad de La Laguna, E-38200 La Laguna, Tenerife, Spain\\
$^{4}$Institute for Astronomy, University of Edinburgh, Royal Observatory, Edinburgh EH9 3HJ, United Kingdom\\
$^{5}$Departamento de F\'isica Te\'{o}rica, M\'{o}dulo 15, Facultad de Ciencias, Universidad Aut\'{o}noma de Madrid, 28049 Madrid, Spain\\
$^{6}$Institute of Astronomy, University of Cambridge, Madingley Road, Cambridge CB3 0HA, UK\\
$^{7}$ Centro de Astrobiolog\'{\i}a (CAB, CSIC-INTA), ESAC Campus, 28692 Villanueva de la Ca\~nada, Madrid, Spain\\
$^{8}$ Kapteyn Astronomical Institute, University of Groningen, Postbus 800, 9700 AV Groningen, The Netherlands\\
}}

\date{Accepted XXX. Received YYY; in original form ZZZ}

\pubyear{2022}

\begin{document}
\label{firstpage}
\pagerange{\pageref{firstpage}--\pageref{lastpage}}
\maketitle
\begin{abstract}
Next-generation wide-field spectroscopic surveys will observe the infall regions around large numbers of galaxy clusters with high sampling rates for the first time. Here we assess the feasibility of extracting the large-scale cosmic web around clusters using forthcoming observations, given realistic observational constraints. We use a sample of 324 hydrodynamic zoom-in simulations of massive galaxy clusters from TheThreeHundred project to create a mock-observational catalogue spanning $5R_{200}$ around 160 analogue clusters. These analogues are matched in mass to the 16 clusters targetted by the forthcoming WEAVE Wide-Field Cluster Survey (WWFCS). We consider the effects of the fibre allocation algorithm on our sampling completeness and find that we successfully allocate targets to 81.7 $\% \pm$ 1.3 of the members in the cluster outskirts. We next test the robustness of the filament extraction algorithm by using a metric, $D_{\text{skel}}$, which quantifies the distance to the filament spine. We find that the median positional offset between reference and recovered filament networks is $D_{\text{skel}} = 0.13 \pm 0.02$ Mpc, much smaller than the typical filament radius of $\sim$ 1 Mpc. Cluster connectivity of the recovered network is not substantially affected. Our findings give confidence that the WWFCS will be able to reliably trace cosmic web filaments in the vicinity around massive clusters, forming the basis of environmental studies into the effects of pre-processing on galaxy evolution.

\end{abstract}





\section{Introduction}

Galaxies spend a significant amount of time in cosmic web filaments. They are the backbone of a complex network of geometrical structures that indicate how matter is distributed in the Universe: 
as filaments, sheets, walls, knots and voids \citep{aragon10}, described by the Zel'dovich formalism \citep{Zeldovich}. 
The present day cosmic web has evolved from small over- and underdensities already existent in the early Universe, and still grows and changes today as a consequence of gravitational collapse. Since $z \approx 2$, about half of the mass budget of the Universe is located within filaments, whilst only contributing to $6 \%$ of its volume. By contrast, only about $10\%$ of the mass is located in galaxy clusters \citep{Cautun14}. 

However, clusters stand out as recognizable, bright peaks in the density field and have been the focus of observational galaxy evolution studies for decades \citep{Gray09, Balogh17}. They have uncovered a now well--supported relation that finds higher fractions of quiescent and early type galaxies in clusters compared to outside of clusters (Morphology-Density relation, \citealt{Dressler97}). Typically, this is explained through astrophysical effects that quench and transform galaxies as they encounter the extremely dense intra-cluster medium of cluster cores during their infall. Ram pressure stripping is one of several possible mechanisms quenching galaxies infalling onto a cluster \citep{Zinger, Arthur19}. However, the majority of the gas lies outside the boundaries where the clusters are virialized, and in the intergalactic medium within filaments \citep{Walker22, Galarraga22, Gouin2022GasIllustrisTNG}. 

Galaxy clusters are therefore not isolated islands, but assemble, replenish and grow via ongoing mergers with smaller clusters, groups and clumps of gas, as well as through a constant flow of gas and galaxies from filaments. The most prominent of these filaments have hot gas temperatures and dense cores that have the possibility to strip the gas from galaxies, but also to replenish galaxies with pre-enriched filamentary gas \citep{vulcani2019}, impacting their mass assembly and star formation histories in very different ways \citep{Laigle18, Song20}. It is clear that the challenge of understanding galaxy evolution must include the impact of the large-scale geometry and flows of the cosmic web, and that galaxy transformation begins well before the galaxies fall into the cluster ("pre-processing", \citealt{Zabludoff_1998}).
Physical processes in the outskirts of galaxy clusters are therefore fundamentally different from cluster cores, and thus important areas for the study of cluster assembly and their connection to the filaments of the cosmic web \citep{Florian,Salerno20, Malavasi22, Gouin2020}. However, they are challenging to capture. 

Whilst filaments can be identified by mapping the gas distribution of galaxy clusters in simulations \citep{Kuchner20, Perez20, Gouin2022GasIllustrisTNG}, galaxies tend to trace these features of the cosmic web and can therefore be used to detect filaments observationally \citep{Einasto20, Malavasi20}. To correctly identify filaments that feed clusters, we require a large area, high sampling density and depth to cover a sufficient number of galaxies over a broad range of masses. Large-area surveys such as the Sloan Digital Sky Survey (SDSS, \citealt{SDSS}), the Galaxy and Mass Assembly survey (GAMA, \citealt{GAMA}), the Two-Degree Field Galaxy Redshift Survey (2dFGRS, \citealt{2degree}), the VIMOS Public Extragalactic Redshift Survey (VIPERS, \citealt{VIPERS}) and the Dark Energy Spectroscopic Instrument (DESI, \citealt{DESI}) all probe the distribution of galaxies over large redshift ranges, providing strong observational evidence for the presence of a cosmic web. However, they either lack statistically significant samples of galaxy clusters, or the necessary sampling or detail required for an investigation on pre-processing by filaments feeding clusters. Targeted spectroscopic studies that focus on clusters may provide the required sampling, but they are either only available as case studies of stand-out targets such as Virgo \citep{Castignani_2021, Castignani_2022}, or do not extend far enough to bridge cluster infall regions to the large-scale cosmic web filaments (e.g., GOGREEN \citealt{Balogh17} or OmegaWINGS \citealt{Gullieuszik_2015, Omegawings}). 

To address the need for observing programmes that combine high sampling and statistical power, we look towards next generation wide-field, multi-object spectroscopic (MOS) surveys of galaxy clusters as they will enable detailed study into the far-reaching lower-density cluster outskirts. They are designed to reveal the complex interplay between the properties of galaxies and their position in the cosmic web filaments that feed the clusters. Examples of next generation MOS surveys are the upcoming WEAVE Wide Field Cluster Survey (WWFCS; Kuchner et al.\ in prep) and the 4MOST CHileAN Cluster galaxy Evolution Survey (CHANCES, Haines et al. in prep). 
We motivate our studies with the WWFCS, which will cover 20 low redshift (z$\sim0.05$) galaxy clusters in a mass range of $\log(M_{\rm cl}/M_{\odot}) = 13.8-15.5$ out to and beyond $5R_{200}$. For each cluster, thousands of new spectra will be obtained with a galaxy stellar mass limit of $10^{9}M_{\odot}$, extending our current understanding of these systems to include the infall regions and low-mass galaxies. 

Given the challenging task of accurately mapping filaments in the vicinity of massive clusters (both in terms of extreme contrasts of interlaced high and low density regions, 2D projections, and complications due to the Finger of God effect \citealt{Kuchner21}), careful preparation is required. We have planned, tested, and fine-tuned our steps to map and characterise the infall regions of these clusters with a large statistical sample of simulated clusters from TheThreeHundred survey, and investigated strategies for doing so in redshift space \citep{Kuchner20,Kuchner21,Kuchner22,Rost21}. To confidently carry out the observational programme with the 1000-fiber fed MOS WEAVE at the Wiliam Herschel Telescope \citep{Dalton14}, we now take the final step from simulations to fully-informative mock observations to understand the success and limitations of identifying filaments around clusters. 

Our goal for this paper is to quantify what effect the physical constraints of assigning fibers to targets---a necessary and important step in the design of a MOS survey---has on filament finding. Our previous investigations assumed that all theoretical cluster structure members are targeted and return spectra, thus featuring in the mapping and subsequent analysis. However, in reality, instruments only have a finite capacity to place fibers on targets, and physical restrictions imposed by the geometry and size of the instrumental components require us to make decisions that will ultimately influence the success of finding filaments. In addition, limitations of a realistic target selection may lead to losing valuable fibers to background galaxies. 
Fiber collisions in dense regions like groups and substructures in the outskirts, as well as decisions on which galaxies should receive higher priority than others, directly link to the input for filament identification algorithms and thus could impact the analysis of pre-processing in infall regions.

In this paper, we therefore close the circle of comparing simulations to observations---from a theoretical 3D volume to a fully configured 2D projection. We design a framework for determining the feasibility of reliably characterizing the large scale structure from galaxies that can be observed in current wide-field cluster surveys, using concrete constraints that are matched to the WEAVE instrument and the WWFCS. 

The paper is structured in the following way: Section 2 describes the data we have used. This includes the spectroscopic survey inspiring this paper and the numerical simulations used to create mock observations. Section 3 reports our generation of mock observations and a summary on how we extract the cosmic web. Section 4 displays the results and discussion, we describe the accuracy in which we trace the large-scale structure surrounding our simulated clusters and explain the importance of our results in probing the success of next-generation spectroscopic surveys. We present our conclusions on the likely success of filament retrieval from WWFCS in Section 5.

\section{Galaxy cluster information}
The framework presented in this paper is designed with the WWFCS in mind, but should also work for similar surveys making the adjustments required by the specific observational strategy and constraints. We will use simulation outputs in tandem with algorithms underpinning the observational processes of next generation wide-field cluster surveys. This section describes the planned observations and simulated data relevant to this work. 

\subsection{The WEAVE Wide-Field Cluster Survey}
\label{sec:WEAVE}
WEAVE (William Herschel Telescope Enhanced Area Velocity Explorer) is a next generation MOS. The spectrograph makes use of $\sim1000$ individual fibres deployable over a 2-degree-diameter field-of-view. The instrument also includes 20 small deployable integral field units (mini-IFUs), as well as one large IFU. In this paper we are only concerned with the (MOS) observing mode. Low and high spectral resolution observing modes are available.  Further details on the instrument can be found in \citealt{Balcells10, Dalton14, Dalton16} and  Jin et al.\ (in prep).

A number of surveys will be carried out with WEAVE in the next few years, one of which is the WEAVE Wide-field Cluster Survey (WWFCS; Kuchner et al.\ in prep). The WWFCS will utilize the 1000 fibre-fed MOS to study the infall regions of galaxy clusters in unprecedented detail. The WWFCS will observe up to 20 clusters at low redshift ($0.04 < z < 0.07$) and will return spectra for thousands of cluster members for each cluster, out to several virial radii. The sample consists of galaxy clusters previously observed in the WINGS \citep{Fasano2006} and OmegaWINGS \citep{Omegawings} surveys. The WINGS sample covers a wide range of cluster masses ($\sigma = 500 - 1200 \text{ kms}^{-1}$; $\log{L_{\text{X}}} = 43.3 - 45 \text{ ergs}^{-1}$; virial masses $\log(M_{\rm cl}/M_{\odot}) = 13.8-15.5$). The WWFCS selected clusters have velocity dispersions and X-ray luminosities that are statistically indistinguishable from the parent sample and are therefore, unbiased in terms of their mass distribution (Kuchner et al.\ in prep). The WWFCS will use the low spectral resolution ($R\sim 5000$) mode and obtain optical spectra in the 366 nm $ < \lambda < $ 959 nm range. These spectra will yield accurate redshifts, velocity dispersions as well as quantitative information on the star formation histories of the different galaxy populations.

\begin{figure}
\centering
    \includegraphics[width = 0.45\textwidth]{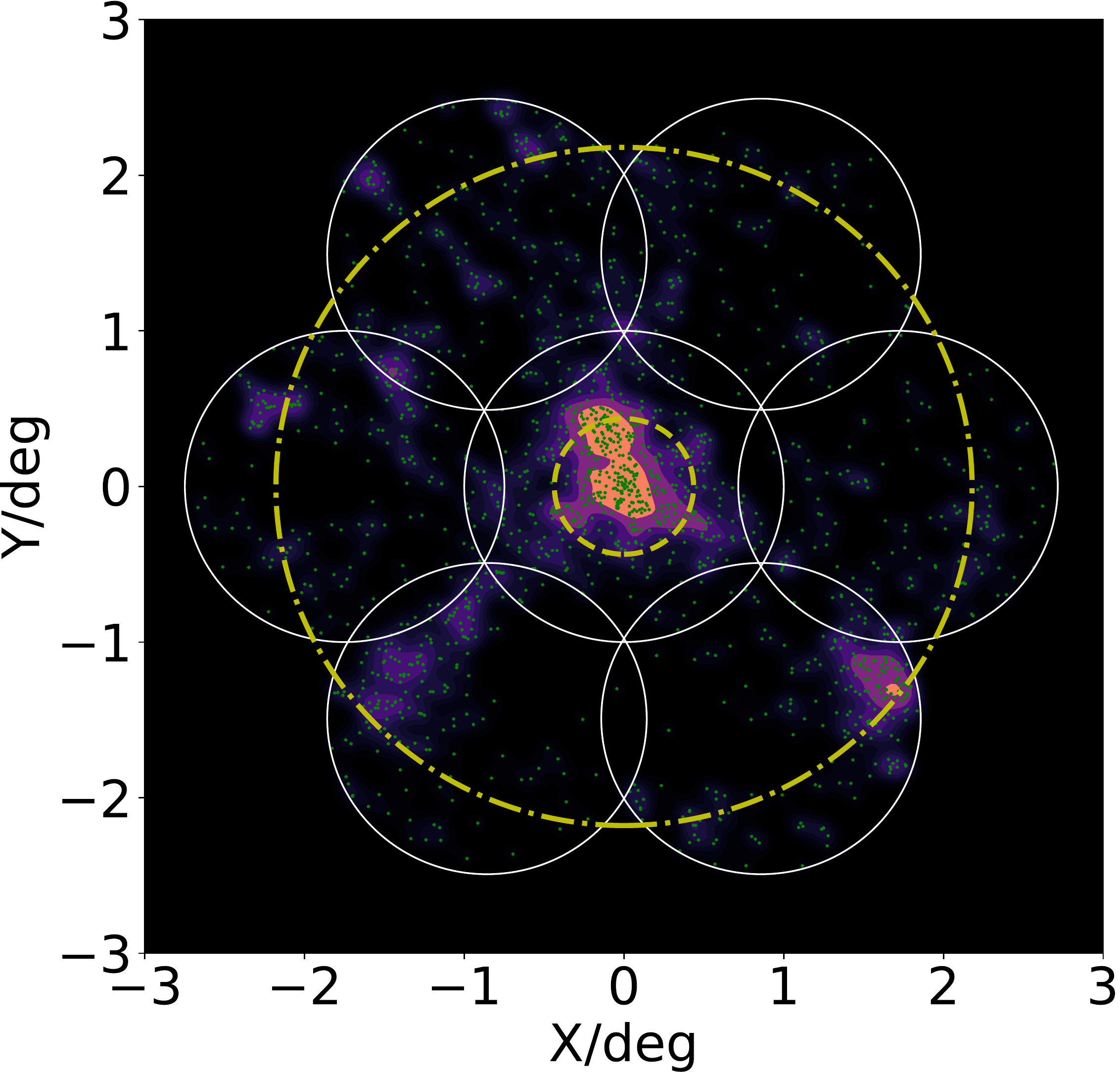}
    \caption{Example simulated cluster from \textsc{TheThreeHundred} with a similar mass and redshift to the cluster Abell 602, one of the WWFCS targets.  The projected dark matter density distribution is shown, derived using a Kernel Density Estimation (KDE) with a~$500\,$kpc smoothing scale. This box has a depth of 10 Mpc. The green dots indicate the positions of galaxy-mass halos. Each white circle encloses a WEAVE field with a 2 degree diameter. The central yellow dashed circle corresponds to $R_{200}$ and the larger dot-dashed yellow circle to $5R_{200}$.}\label{fig:fig1}
\end{figure}

The WWFCS observing strategy is illustrated by Figure~\ref{fig:fig1}. We show a simulated galaxy cluster (cf. Section \ref{sec:The300}) overlaid with WEAVE 2-degree diameter MOS fields (white circles). The inner yellow dashed circle corresponds to the cluster's $R_{200}$, the radius at which the density is equal to two hundred times the critical density of the Universe. The outer yellow dot-dashed circle corresponds to $5R_{200}$. Note the large over- and under-dense regions reaching far out from the very dense cluster core. The large field of view we will be able to cover with WEAVE will allow us to explore and map these environments -- including filaments -- in great detail, reaching much larger cluster-centric distances than hitherto possible (beyond $5R_{200}$), and also study the properties of the galaxies that inhabit them. 
\\

\subsection{\textsc{TheThreeHundred} galaxy cluster simulations}
\label{sec:The300}
\textsc{TheThreeHundred}\footnote{\url{https://the300-project.org/}} project \citep{Cui2018} is a set of zoom-in resimulations of the Multidark Dark Matter only (MDPL2) cosmological simulations \citep{Klypin2016}. MDPL2 is a periodic cube of comoving length $1\,h^{-1}\,$Gpc containing $3840^3$ dark matter particles, each with mass $1.5 \times 10^{9}\,h^{-1} {M_{\odot}}$. MDPL2 uses \textit{Planck} cosmology ($\Omega_{\text{M}} = 0.307,  \Omega_{\text{B}} = 0.048,  \Omega_{\Lambda} = 0.693,  h = 0.678,  \sigma_8 = 0.823,  n_s = 0.96$). 

This simulation suite extracts the 324 most massive individual regions ($M_{\text{vir}} > 8 \times 10^{14} h^{-1}M_{\odot}$), follows them back to their initial conditions and resimulates the hydrodynamics of the volume surrounding a 15 $h^{-1}$ Mpc radius sphere enclosing the cluster and its environment at a higher resolution. Outside of this high resolution region are a set of consecutive shells, hosting lower mass resolution particles that reproduce the tidal fields of the large-scale structure at a reduced computational cost. The highest resolution dark matter particles are divided into dark matter and gas, following the cosmological baryonic fraction using the Planck 2015 cosmology: ${\Omega_{\text{b}}}/{\Omega_{\text{M}}} \approx 0.16$. This gives a combined mass resolution of $m_{\text{DM}} + m_{\text{gas}} = 1.5 \times 10^{9} h^{-1} M_{\odot}$. There are 128 individual time snapshots for all 324 zoomed-in Lagrangian regions, ranging from $z = 17$ to $z=0$. Throughout this work, we use the last snapshot at $z = 0$ given the low redshift of the WWFCS sample. 

These zoom-in re-simulations have been run with the smoothed-particle hydrodynamics codes (SPH): \textsc{Gadget-MUSIC} \citep{Sembolini13}, \textsc{Gadget-X} \citep{Beck2015, Rasia_2015} and mesh-less code \textsc{GIZMO-SIMBA} \citep{Dave19, Cui22}. We only focus on \textsc{Gadget-X} which incorporates full-physics galaxy formation, star formation and feedback from both SNe and AGN. The work in this paper utilizes the AMIGA Halo Finder \citep{Gill2004,Knollmann2009} to determine the halo properties. 

\textsc{TheThreeHundred} simulations provide a useful testbed to develop the observational strategy and forecast the performance of the WWFCS. First, the large volume of the parent dark-matter simulation (MDPL2) ensures a high number of massive clusters are available for statistical purposes. Second, the high-resolution re-simulations reach out as far as $15 h^{-1}\,$Mpc from each cluster centre, comparable to the area that the WEAVE observations will cover. The extensive information available from the cluster centre all the way to beyond $5R_{200}$ allows us to study all the environments present -- from individual galaxy halos to filaments, groups, and the cluster core.  

This dataset has already been used previously to generate theoretical expectations with the WWFCS in mind. For instance, \cite{Kuchner21} quantified the impact of redshift space distortions ('Fingers of God') on the identification of cosmic filaments. They found that trying to correct for this effect statistically in the virialized regions of clusters and groups does not lead to a more reliable extraction of the `true' filamentary networks. For this reason, \cite{Kuchner21} forecast that the identification of the cosmic web in the regions surrounding massive clusters using spectroscopic surveys should rely primarily on the 2D positions of the galaxies on the sky. However, they also point out that accurate spectroscopic redshifts are crucial in defining and isolating the cluster volume from which these galaxies should be selected.

\section{Generation of mock observations}
This section describes the framework we have developed to create mock observations and their optimization using the simulated clusters. We also discuss how we use WEAVE's fibre allocation algorithm to generate realistic WWFCS-like `simul-observed' galaxy samples from the simulations, and the method for identifying the cosmic web using these samples. In other words, we describe the steps we take to go from simulations to observations.

\subsection{Optimizing the WWFCS field positions}
\label{sec:fields}
It is important to optimize the observational strategy of upcoming wide-field spectroscopic surveys to improve the reliability of the filament extraction process while maximising the observational efficiency. The WWFCS can place MOS fibres on targets over a 2-degree-diameter field (Figure~\ref{fig:fig1}). In order to map the filamentary structures that feed clusters it is necessary to maximise the spatial coverage, reaching out to and beyond $5R_{200}$. Such radial coverage is a good compromise between the available observing time and the need to cover as far as possible into the infall regions around the clusters (Kuchner et al.\,, in prep.). We therefore need to design a tiling strategy to cover a circular region around the clusters that reaches $5R_{200}$ using the minimum number of WEAVE fields (or pointings). The tiling strategy we have used to find the optimal position of the WEAVE pointings for each WWFCS cluster is described in detail in Appendix ~\ref{sec:A1} (see examples in Figures~\ref{fig:fig1}, \ref{fig:fig4}, and~\ref{fig:fig5}). The `simul-observations' described below follow the same strategy. We note that by applying this optimisation process we have been able to reduce the required number of pointings (and thus the required observing time) by $\sim 15\%$ from our initial estimate, allowing us to increase the number of clusters we will be able to observe in the available time from $\sim16$ to $\sim17$--$19$ without compromising the accuracy of our filament mapping.

\subsection{Deriving WWFCS cluster properties}
In order to develop mock-observations from the simulations, we need to determine the properties of the clusters selected for the WWFCS. We firstly calculate $R_{200}$ and $M_{200}$ (the radius and mass where the density is 200 times the critical density of the Universe) of the clusters using their spectroscopic redshift $z$ and velocity dispersion $\sigma$ from the WINGS survey \citep{Omegawings} using the following equation \citep{Poggianti2010,Finn}:
\begin{equation}
    R_{200} = \frac{1.73\sigma}{(\Omega_{\Lambda} + \Omega_{\text{M}}(1+z)^{3})^{1/2}}.
\end{equation}
The cluster mass inside $R_{200}$ ($M_{200}$) can then be estimated using this value and the critical density of the universe. The complete list of the WWFCS targets and their properties can be found in Kuchner et al. (in prep.).
The bottom panel of Figure~\ref{fig:fig3} shows the mass distributions of the WWFCS cluster sample (blue) and those from \textsc{TheThreeHundred} simulations (orange). The mass distribution of the simulated clusters is skewed towards higher masses than those of the clusters selected for the WWFCS. 
This is to be expected since \textsc{TheThreeHundred} resimulates the most massive haloes in a large cosmological volume, making it possible to find the rarest objects. By contrast, the clusters selected for the WWFCS are more representative of clusters at low redshift and deliberately span a large range in X-ray luminosity (Section \ref{sec:WEAVE}). We address this mismatch in next section. 

\subsection{Generating the simulated cluster and galaxy samples}
\label{sec:data_300}
We discuss now the generation of the clusters and galaxies that will be included in our mock observations using the halo data from \textsc{TheThreeHundred}. The main aim is to create mock sample analogues to the ones we expect from the WWFCS.

First, we impose some quality constraints on the cluster halos in the simulations so that we only select the highest quality data. We confine our study to the high resolution region of the cluster re-simulation, a spherical region with a radius of $15\,h^{-1}\,$Mpc centered on the cluster centre. We then require that the mass fraction in high resolution particles for the zoom simulation needs to be greater than 0.99 ($fM_{\text{hires}} > 0.99$). This criterion rejects low-resolution dark matter particles that may have travelled inwards into the high resolution region during the simulation of the clusters evolution. Finally, we only accept halos with a mass greater than the simulation resolution ($3 \times 10^{10} M_{\odot}$), corresponding to 20 dark matter particles, as explained in \citealt{Kuchner20}. 

\subsubsection{Creating mass-matched cluster samples}
\label{sec:mass_match}
In order to make a meaningful comparison between the clusters from \textsc{TheThreeHundred} simulations and the WWFCS clusters, we need to create a sample of simulated clusters whose masses match those of the observational sample. Past studies have shown that for a flat Universe, on scales large enough to neglect baryonic physics, dark matter halos evolve self similarly \citep{Kaiser86,Mosotoghiu19}. Self-similarity implies that the dark matter distribution (and hence the location of dark-matter halos) in less massive galaxy clusters is well represented by that of more massive clusters that have been scaled down appropriately taking into account their mass ratio. 

To have reasonable statistics, our goal is to create a mass-matched sample of simulated WWFCS cluster analogues containing 10 simulated clusters for each WWFCS cluster.  Because dark matter halos evolve self similarly on large scales, we are able to do so using the large sample of simulated clusters from \textsc{TheThreeHundred} project. 

We describe in detail the methodology behind the mass-scaling of the simulated clusters in Appendix \ref{sec:A2}. In short, when the mass of a WWFCS cluster is too small to be able to find similar mass clusters in \textsc{TheThreeHundred} simulations, we use a mass scaling factor $M_{\text{F}}$ to reduce the mass of all the simulated clusters to approximately match the mass of the WWFCS cluster. The mass of each individual dark matter halo in the relevant cluster simulation is also scaled down by the same factor, and included in the halo sample if its mass is above a mass threshold. These mass thresholds are chosen to ensure that the number of galaxy-mass halos in each simulated cluster approximately matches that expected in the the observed cluster. For each WWFCS cluster, ten analogue clusters with similar masses are randomly selected from the scaled clusters.

\begin{figure*}
    \centering
    \includegraphics[width = 0.85\textwidth]{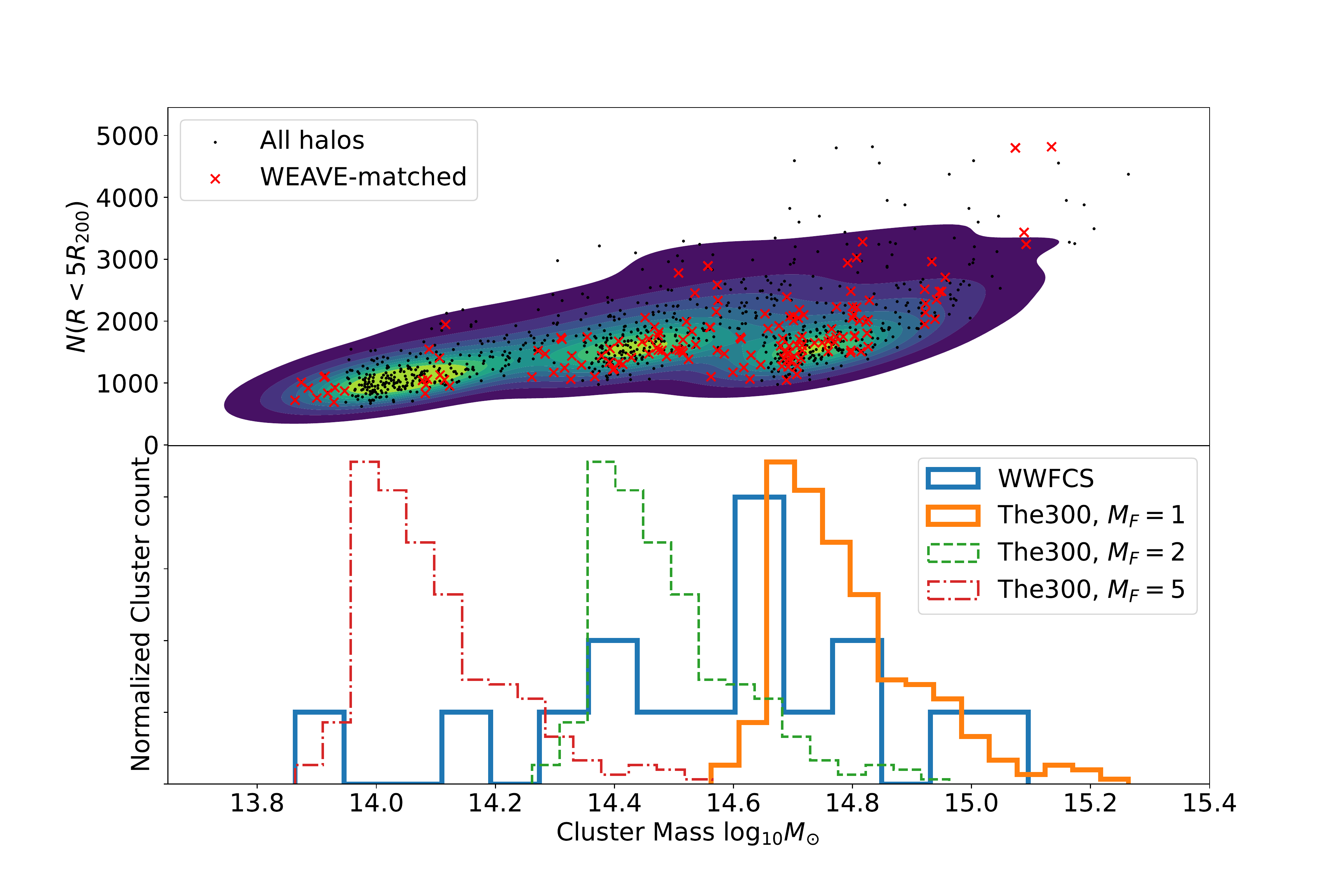}
    \caption{Bottom panel: normalized cluster count as a function of cluster mass showing the mass distribution of the WWFCS clusters (solid blue line) and \textsc{TheThreeHundred} un-scaled clusters ($M_{\text{F}} = 1$, solid orange line), and scaled by $M_{\text{F}} = 2$ (green dashed line) and $M_{\text{F}} = 5$ (red dot-dashed line). Three scaling values ($M_{\text{F}} = 1, 2,$ and~$5$) are sufficient to fully cover the mass range of the WWFCS cluster sample. Top panel: the number of galaxy halos contained within $5R_{200,\text{scaled}}$ for each simulated cluster as a function of cluster mass is displayed by the small dots and the coloured density distribution. Red crosses correspond to the mass-matched simulated clusters (ten per WWFCS cluster). This gives an indication of the approximate number of galaxies that can be `observed' in each cluster simul-observation. }
    \label{fig:fig3}
\end{figure*}

Figure~\ref{fig:fig3} demonstrates that three values of the mass-scaling factor $M_{\text{F}}$ (1, 2 and 5) are sufficient to provide enough mass-matched simulated clusters. Of course, $M_{\text{F}}=1$ implies no scaling is applied. The bottom panel shows the mass distribution of the WWFCS clusters (black) together with that of the un-scaled and scaled \textsc{TheThreeHundred} clusters. The orange, green and red histograms represent the mass distributions of all~324 simulated clusters with $M_{{F}} = 1, 2,$ and~$5$ respectively.

\subsubsection{Scaling cluster properties}
Having built a mass-matched sample of simulated clusters, we now describe how the spatial properties of the clusters and subhaloes (linear and angular size, halo positions) are likewise scaled.

The radius of the mass-scaled clusters $R_{200,\text{scaled}}$ can be derived from the relationship between $R_{200}$ and $M_{200}$, adapted from \citealt{Poggianti2010}, 
\begin{equation}
    R_{200}^{3} = \frac{M_{200}}{K \times h^{2}(\Omega_{\Lambda} +  (1+z)^{3}\Omega_{0})},
\end{equation}
where $K = 2.32 \times 10^{14}\,M_\odot$Mpc$^{-3}$, making $M_{200} = M_{200,\text{scaled}}$. The spatial coordinates $x$, $y$ and $z$ of all the halos in the cluster (and thus their clustercentric distances) are therefore multiplied by a factor $R_{200,\text{scaled}}/R_{200, \text{unscaled}}$. 

For each WWFCS cluster the angular diameter distance is calculated using their redshifts \citep{Omegawings} with the adopted cosmology. Next, for each of the corresponding ten analogue simulated clusters, we convert 3D positions to angular distances between the haloes and the centre of their clusters, and thus their relative positions on the simulated sky.  

The top panel of Figure~\ref{fig:fig3} displays the number of galaxy halos contained within $5R_{200,\text{scaled}}$ from their cluster centers as a function of cluster mass (small dots and density distribution). The red crosses correspond to the clusters in the mass-matched sample (ten per WWFCS cluster). Note that the planned WWFCS observations will generally cover well beyond~$\sim5R_{200}$ (cf. Kuchner et al., in prep.; see also~Figure~\ref{fig:fig1} and Appendix~\ref{sec:A1}), and therefore the number of potential targets for each cluster shown in the figure is a conservative lower limit.

\subsubsection{Defining galaxy sample and properties}
\label{sec:galprop}

To bring our mock `observational' sample closer to the real observations, each of the WWFCS clusters' mass-matched simulated analogues are placed at the appropriate redshift and sky location. We then allocate WEAVE pointings using the field positions determined in Appendix \ref{sec:A1}. Only halos covered by these pointings will be considered as possible spectroscopic targets. 

Galaxy-size dark-matter halos in each simulated cluster are then given in-fibre magnitudes in the SDSS $r$-band (similar to the ones that will be used the observational target selection) following a simple procedure that ensures the target galaxies have comparable numbers and magnitude ranges to the planned observations. The actual galaxy magnitudes have no impact on the findings of this paper, but the fibre allocation program \texttt{Configure} \citep{Terret14} requires them as input. Explicitly, the total $r$-band magnitude of a galaxy is estimated from the mass of the simulated halo using the equation
\begin{equation}
    r_\text{total} = W - 2.5\log_{10}({M_\text{halo}/M_{\odot}}),
\end{equation}
where $W$ is a constant that is calculated by mapping the least massive halos in each simulated cluster (Appendix \ref{sec:A2}) to the planned $r$-band limit of the WWFCS spectroscopic observation ($r_\text{total} < 19.75$, corresponding to an approximate galaxy stellar mass limit of~$\sim10^9\,M_\odot$, Kuchner et al., in prep). An average offset between total and in-fibre magnitudes of $1.75\,$mag, estimated through a least-square fit to the appropriate SDSS magnitudes, is then applied. The in-fibre magnitude limit of the WWFCS galaxy sample is therefore $r_\text{fibre} < 21.50$, which sets the planned exposure times of $\sim1\,$hour. This exposure time is expected to yield reasonable signal-to-noise ($S/N > 5$ per \AA, for all the spectra up to this magnitude limit), and we therefore expect close to 100\% redshift completeness for the observed (and thus mock) galaxies (see Kuchner et al. in prep.).We use a simple procedure to allocate magnitudes to the galaxy-sized dark matter halos here, as the results in this paper only require accurate spatial distributions of mock galaxies and their expected number densities. Our simple approach ensures this without relying on uncertain model galaxy properties. As  \cite{Cui2018} show (see, e.g., their Figure~8), large uncertainties still remain in the model observed magnitudes and colours, and the results depend strongly on the specific baryonic model used, particularly at low galaxy masses. While the simulations have appropriate resolution to yield reliable masses and locations for the dark matter haloes, the additional step of predicting observable properties through the available hydrodynamic or semi-analytic models would require making uncertain choices which are not necessary for our purposes.

At this point we have created a set of 160 simulated galaxy clusters (10 per WWFCS target cluster), populated them with mock galaxies, placed them at the appropriate redshift and sky position, and covered them with WEAVE pointings exactly as those planned in the observations. 

\subsection{Allocating spectroscopic fibres to mock galaxies using \texttt{Configure}}
\label{sec:config}
An integral part of creating mock observations is to carry out a realistic  allocation of spectroscopic fibres to the mock galaxies since, this process can potentially distort and limit the spatial information that can be derived from the real observations. Geometric and mechanical constraints (such as fibre collisions and overlap) mean that it is not possible to assign fibres to all the galaxies on a pointing. 

Optimising fibre allocation is not a trivial task, and sophisticated software is generally used to reduce costly human intervention. \texttt{Configure} is the program that WEAVE will use to find an optimal set of assignments of fibres to positions on the sky \citep{Terret14}. Each field (or pointing) will contain not only science targets, but also a set of calibration objects and guide stars.  \texttt{Configure} uses a probabilistic technique named `simulated annealing' \citep{annealing} to simulate the thermal motion of a system to be optimized. The `energy' of each fibre with a target assigned to it is given by $(1.0+s)/p$, where~$s$ is a measure of how straight the fibre is and~$p$, the target priority, is an integer value between 0 and 10 that is used to prioritize objects on the fields. In our case, we assign a maximum priority of 10 to all of the cluster mock galaxy members, and lower values to other targets (Table~\ref{tab:priorities}). The algorithm then optimises the fibre allocation by finding the configuration with the lowest `energy' by swapping the position of fibres until the minimum is found. 

This process determines the galaxies that will be allocated a fibre and therefore decides which galaxies will have spectroscopic information.  It may play a crucial role in determining the accuracy in extracting cosmic-web information from the WWFCS observations, and its effect will be thoroughly tested in this paper. 

\begin{figure*}
    \centering
    \includegraphics[width = \textwidth]{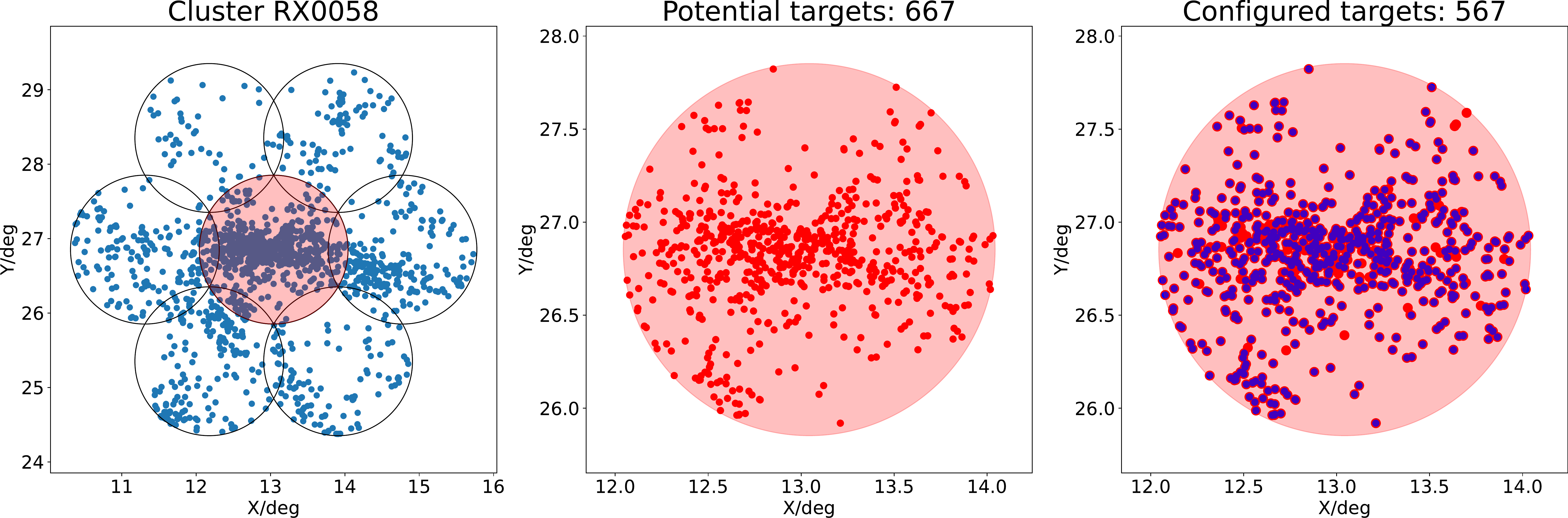}
    \caption{A demonstration of the process of `configuration', the allocation of spectroscopic fibres to targets, on a simulated cluster mass-matched to WWFCS cluster RX0058 ($M\sim4.3 \times 10^{14} M_{\odot}, R_{200} \sim 1.54\,$Mpc), an average mass WWFCS cluster. Left panel: mock `observation' of the simulated cluster containing~8 2-degree diameter WEAVE fields. Blue points show the positions on the sky of the simulated galaxies. Middle panel: zoom-in on the central field from the plot on the left, with the simulated target galaxies before fibre configuration shown as red points. There are 667 simulated cluster members that are potentially observable in this field. Right panel: the same central field after configuration, where 567 cluster members have been assigned a fibre (blue dots), while galaxies without a fibre are shown in red.}
    \label{fig:fig4}
\end{figure*}

\begin{figure*}
    \centering
    \includegraphics[width = \textwidth]{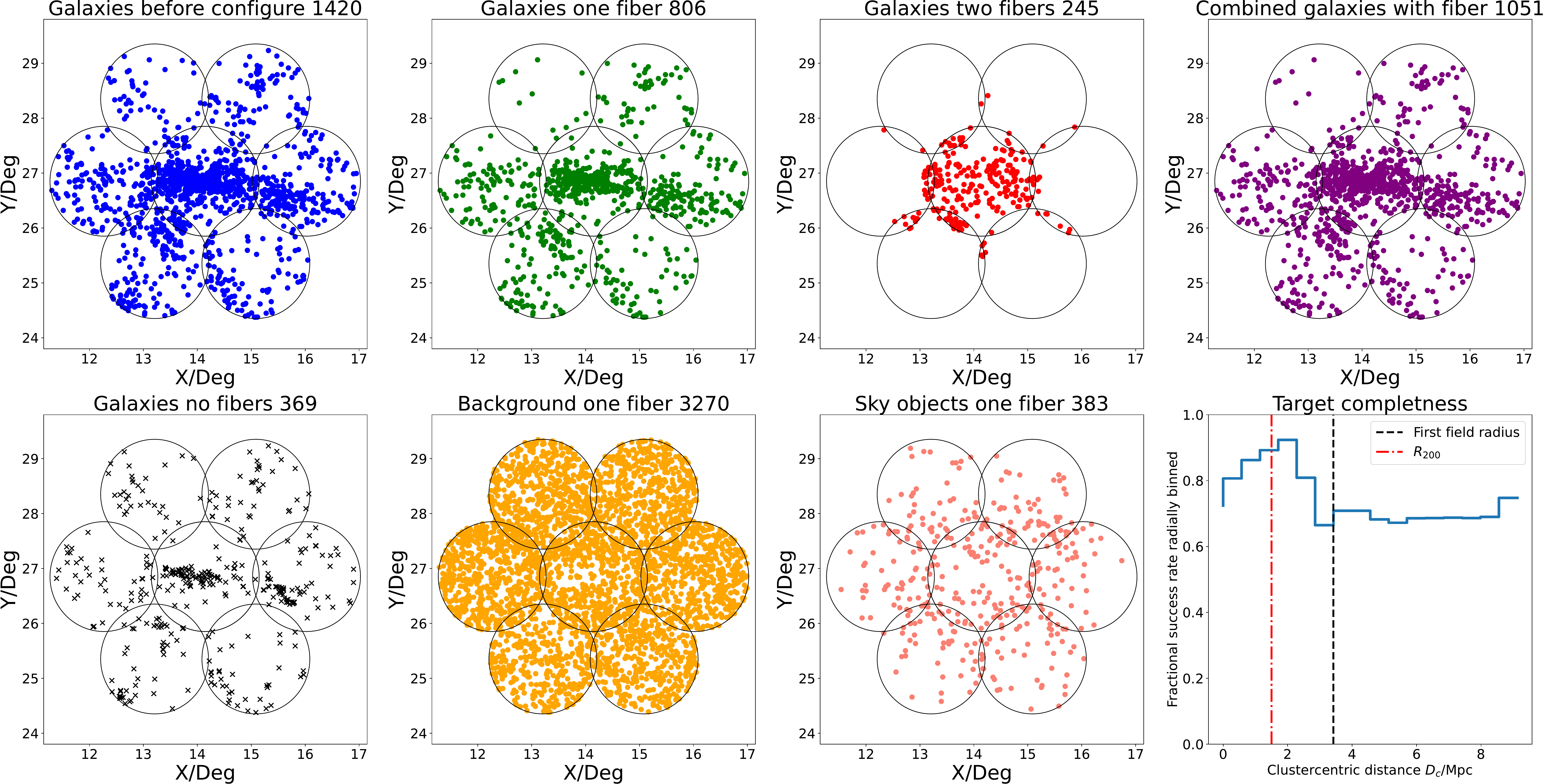}
    \caption{Results for the full configuration process of one simulated analogue to cluster RX0058. We plan to observe this cluster using 8 WEAVE individual pointings (black circles). The central pointing will be observed twice given the high density of targets in the cluster core. For this cluster, $74\%$ of its simulated galaxies have at least one fibre allocated and $69\%$ background objects also have allocated fibres. Top left: all cluster members (blue dots) that could be assigned a fibre. Top middle-left: galaxies with one fibre assigned (green dots). Top middle-right: galaxies with two fibres assigned (red dots). Top right: combined sample of target galaxies with one or two fibres (purple dots) Bottom left: galaxies with no fibre allocated (grey crosses). Bottom middle-left: background objects allocated one fibre (orange dots) and blank sky positions assigned one fibre (salmon dots). Bottom right: fractional target completeness (i.e., fraction of cluster galaxy targets with one or two fibres; see Section~
    \ref{sec:memberrecoveryrate}) in clustercentric distance bins. $R_{200}$ and the radius of the central WEAVE field are plotted for reference. The bin width is 1/6 of of the radius of one WEAVE pointing (i.e., 1/6 of one degree).}
    \label{fig:fig5}
\end{figure*}

To make a more realistic analogue to the planned observations, before running \texttt{Configure} we pollute our target catalogues with background objects. We randomly place 1400 background objects on each WEAVE field, corresponding to a number density $\sim450\,$deg$^{-2}$. This is somewhat larger than the galaxy number density corresponding to an in-fibre magnitude limit $r_\text{fibre}<21.5$, the planned WWFCS limit, but over-populating the background is not a problem because these objects will be assigned a low priority of~$1$ (see Table~\ref{tab:priorities}). We note that the background objects are not designed to be representative of the larger-scale cosmic web, but are implemented to test the usage of free fibers (i.e., fibers that haven't been assigned to cluster members according to their photometric redshifts). This is especially relevant further away from the cluster center as fibers are then free to be placed on background targets if they are brighter than the magnitude limit. In this paper, we stress-test this assumption by putting a slightly exaggerated number of background targets in the catalog to compete with higher priority cluster targets. If their spectroscopic redshifts from the WWFCS reject them from cluster membership, they will not feature in the filament finding, as described in Section \ref{sec:disperse}. For this paper, we assume that the vast majority of our WEAVE target selection---based on magnitude, colour and precise photometric redshifts---correctly rejects galaxies that lie outside the volume in space that  corresponds to \texttt{TheThreeHundered} volume. This is supported by our tests using available observational data of the cluster centers. While we will only know the exact number  after analysing early WEAVE observations, we can expect that not every background galaxy will be identified correctly. In practice, we will therefore use a range of priorities for background objects. However, given the high quality of the J-PLUS photometric redshifts \citep{Cenarro19}, in combination with conservative colour and magnitude cuts, we expect the percentage of interlopers to be small and that the majority of background objects can be accurately de-prioritised. We keep them in the target catalog for the sole reason that unallocated fibres can be used and they do not feature in our filament finding, as described in Section \ref{sec:disperse}.

Finally, the input catalogues fed to \texttt{Configure} contains also blank sky positions for sky-subtraction purposes. These correspond to real celestial positions devoid of objects visible in SDSS images of the target clusters. They are given a priority of~1. Although in all cases we have more suitable sky positions, \texttt{Configure} is set to allocate a maximum of~50 sky fibres per field, as per the observational strategy described in Kuchner et al. 

Therefore, each WEAVE field target list consists of $N_{\text{gal}}$ cluster galaxies, determined by the simulated cluster galaxy sample (see Section~\ref{sec:galprop} and Figure~\ref{fig:fig3}), $N_{\text{sky}}$ sky positions, and $N_{\text{back}} = 1400 - N_{\text{gal}} - N_{\text{sky}}$ background targets. As mentioned above, the exact number of background objects does not matter, and we limit $N_{\text{back}}$ in this way to keep the size of the target catalogues small enough to keep the \texttt{Configure} running time manageable.

To fully `configure' a cluster's mock-observation we need to apply the \texttt{Configure} software to each individual WEAVE field (or pointing) sequentially, taking into account that these pointing overlap (Figures~\ref{fig:fig4} and~\ref{fig:fig5}) and that the central pointing will be observed twice in order to deal with the high density of targets in the cluster core (Kuchner et al., in prep.). The aim of the process is to maximise the number of target galaxies with at least one fibre allocated. Maximizing the number of galaxies with measured spectroscopic redshifts, particularly in the cluster outskirts and infall regions, is a key goal of the observational strategy that will enable a more accurate mapping of the cosmic web.

Each field intersects a minimum of 3 other fields, meaning that target objects in the overlap region have multiple chances of having a fibre allocated. To obtain information on data quality and repeatability, it is desirable to have some repeated observations, but we do not want these to have a significant impact on the final sample of galaxies with redshifts. We therefore aim at no more than $\sim10$--$20\%$ of the cluster galaxies to be observed twice and we chose to not artificially select an upper limit on the number of galaxies observed twice as we want to minimize the number of empty fibers. We thus allow target galaxies to have at most two fibres allocated (in separate pointings), but sky positions and background galaxies are only allocated one fibre at most. This process is controlled by the \texttt{Configure} targeting priorities (Table~\ref{tab:priorities})\footnote{In this exercise we do not include the flux calibration targets and guide stars since given their small numbers they have a negligible effect in our results.}.

 For all simulated clusters we sequentially apply  \texttt{Configure} to each WEAVE field (see Figure~\ref{fig:fig5}). We start with the central one, which we configure twice, and then continue with the outer fields. After each step we update the priorities for all objects in the target list taking into account whether an object (cluster galaxy, background galaxy or sky position) has been allocated a fibre in a previous iteration. If a cluster galaxy has already been allocated one fibre, its priority is reduced to~1. If it has already been allocated~2 fibres, its priority goes to~0. Background galaxies and sky positions with fibres allocated previously get also a priority of~0 (Table~\ref{tab:priorities}). The process is illustrated with one example for the cluster RX0058 in Figures~\ref{fig:fig4} and~\ref{fig:fig5}. Obviously we are not able to allocate a fibre to each target cluster galaxy, In the typical cluster shown in these figures, $74\%$ of the cluster galaxies have at least one fibre assigned, with little radial variation beyond the radius of the inner WEAVE field. The success rate there is higher despite the higher density because this field is observed twice. Beyond that, no strong spatial biases are apparent, but we will analyse quantitatively the effect the configuration process has on our ability to map the large-scale structure and filaments around the clusters next.

\begin{table}
\caption{Target priorities used in \texttt{Configure}. }
\label{tab:priorities}
\begin{tabular}{lcc}
\hline
Object type & Target priority $p$\\
\hline
Cluster galaxy & 10 \\
Background galaxy & 1\\
Sky position & 1\\
Cluster galaxy with one fibre already allocated & 1\\
Background galaxy with one fibre already allocated & 0\\
Sky position with one fibre already allocated & 0\\
\hline
\end{tabular}
\end{table}

\subsection{Cosmic web extraction method}
The rationale of this paper is to assess the ability of upcoming spectroscopic surveys such as the WWFCS to accurately map and characterise the cosmic web. We describe in this section the techniques we use for that purpose. 

\subsubsection{DisPerSE}
\label{sec:disperse}
To map the large-scale structure around clusters and extract filaments, we utilize the topological structures extractor \texttt{DisPerSE} \citep{Sousbie11,Sousbie_2_11}, which uses the concept of Morse theory \citep{Morse} applied to matter distributions. It identifies structures, such as nodes, walls, knots, and voids as components of the Morse-Smale complex -- the set of all ascending manifolds of the input function -- and is able to classify regions using critical points and integral lines (nulls of the density field and tangents to the critical points). \texttt{DisPerSE} identifies topologically significant regions in the Delaunay-2D/3D tessellation density field by taking a set of discrete 2D or 3D positions (positions of haloes, mock galaxies, gas particles, etc.).  
It computes a set of segments representing the skeleton of the filamentary network. 

In what follows, \texttt{DisPerSE} is run in 2D on the sky positions of the simulated cluster galaxies to mimic the observations since, uncertainties in the radial position of the galaxies due to peculiar velocities mean than filament extraction in 2D is preferable when redshifts (and not true distances) are available \citep{Kuchner20,Kuchner21}.

\begin{figure}
    \centering
    \includegraphics[width = 0.5\textwidth]{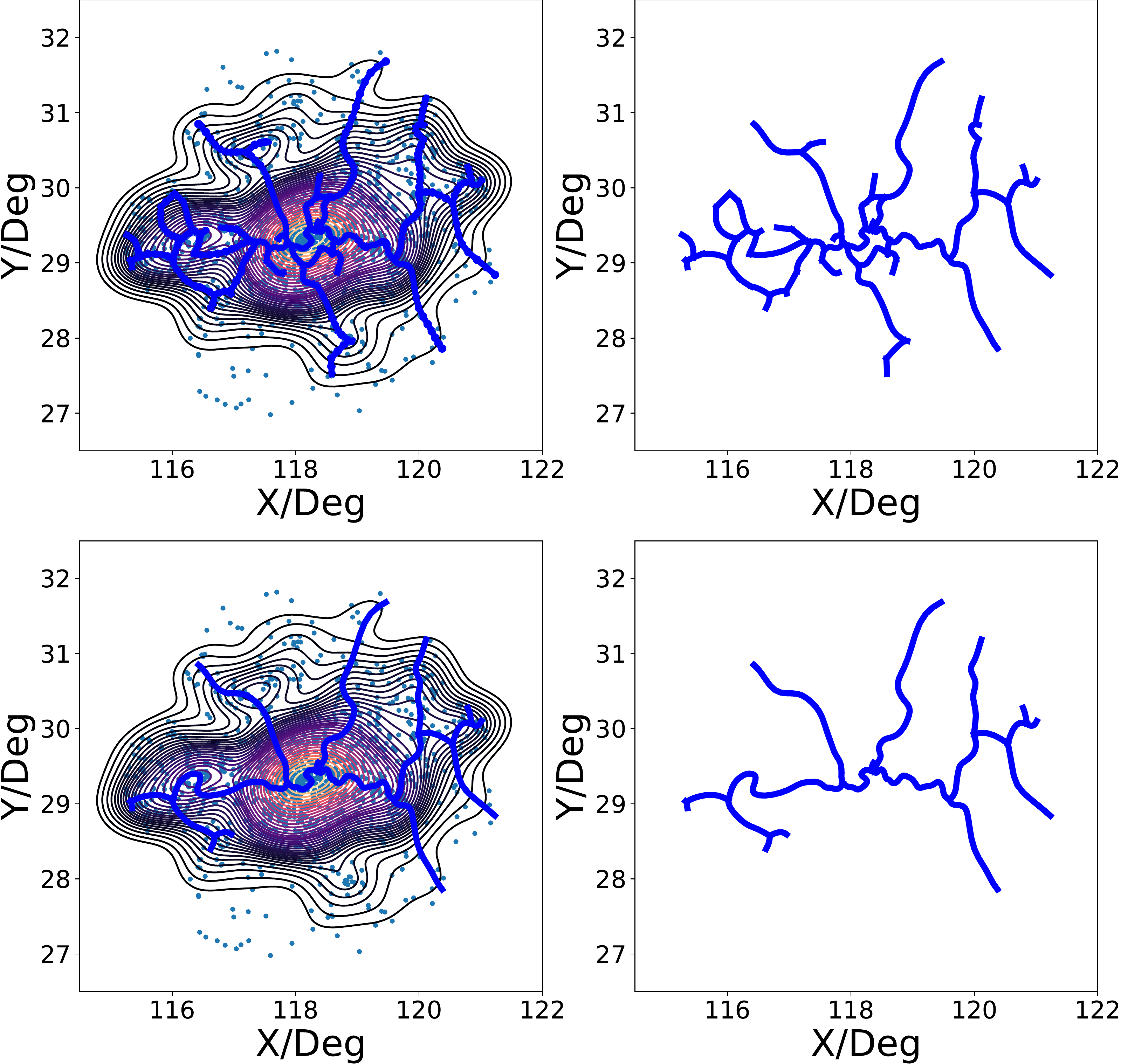}
    \caption{Example filament network extracted with different persistence thresholds. Top row: $\text{persistence} = 2\sigma$. Bottom row: $\text{persistence} = 3\sigma$. All plots correspond to the same cluster. Left column: KDE smoothed matter density distribution with the extracted filament network overlaid. Right column: filament network only.}
    \label{fig:fig6}
\end{figure}

\texttt{DisPerSE} analyses the topology of the pre-defined region and finds the saddle points that join together the main nodes set by a pre-defined `persistence' threshold. The theory of persistence allows a user to account for uncertainty and Poisson noise in data sets (in terms of $\sigma$) and is analogous to signal-to-noise ratio for observations. When extracting filaments with \texttt{DisPerSE}, setting a higher persistence threshold returns only the most robust, topologically significant, large scale structure. Lower persistence values enable the detection of smaller tendrils. Therefore, there is a trade-off between the number of filaments that are extracted and their astrophysical significance. Figure~\ref{fig:fig6} demonstrates the implementation of different persistence thresholds. Both filamentary networks shown are derived by \texttt{DisPerSE} with the same input galaxy positions but using two different persistence values, $2\sigma$ and~$3\sigma$. It is evident that the least robust filament segments are not present in the network associated with higher persistence. Therefore, persistence strongly dictates the level of structure that is extracted, and it is important the its value is optimised for the scientific application intended. In the analysis that follows we will use a persistence of $2.5\sigma$ for the simulated reference network and $2.1\sigma$ for the network obtained from the analogue observations (after \texttt{Configure} is applied). We use different persistence values as the underlying density field will change upon target selection for each cluster. These choices are justified in Appendix~C. There is also a smoothing parameter that is input into the Disperse runs that influences the rigidity of the identified filament paths. We chose a smoothing parameter of 5, as used in \cite{Kuchner20}, although using values between 1 and 5 has virtually no impact our results.

\subsubsection{Dealing with boundary conditions}
The complex 2D geometrical shape defined by the positions of the WEAVE pointings that will tile each cluster and its environment (Appendix \ref{sec:A1}) influences the features that are detected by \texttt{DisPerSE}, particularly near the boundaries. If the shape of the field tiling is not properly accounted for,  artificial nodes are detected that trace the outer boundary of the sky region covered by the WEAVE fields. To avoid that, we populate the region outside the boundaries of the area covered by the planned WEAVE pontings with a random uniform distribution of artificial galaxies that will act as `guard particles' \citep[cf.][]{Sousbie11} to prevent the appearance of these artificial nodes and their associated filaments. The surface density of the artificial galaxies is chosen to be similar to that of cluster galaxies in the outer regions. In practice, the number density of galaxies that lie beyond $2R_{200}$ is computed for each cluster, and a random uniform distribution of `guard particles' with this number density is added outside the outer boundary of the `observed' fields, reaching $7.5R_{200}$. After testing different values for this radius, we find that the recovered networks are very similar when one increases the guard particle boundary beyond $7.5R_{200}$. This is sufficiently far away from the cluster centre to prevent the true filamentary network being distorted by the irregular boundaries. 

The positions of cluster members and 'guard particles' are fed into \texttt{DisPerSE} and, once the filament network is computed by the filament finder, we truncate the network outside of the `observed' region, keeping only the filament segments inside. This procedure works remarkably well, and visual inspection indicates that spurious nodes and filaments associated with the boundaries are eliminated without affecting the filament network inside the observed fields.

\section{Results and discussion}
With all the necessary elements in place, in what follows we will compare the filamentary networks that are extracted using \texttt{DisPerSE} before and after applying the MOS `fibre configuration' process. In other words, we will quantify the difference between the filaments extracted when all simulated galaxies are fed into \texttt{DisPerSE} with those we obtained if we only use the `mock-observed' galaxy sample, where some galaxies are lost due to fibre-positioning constraints. This will allow us to forecast the impact that realistic observational constraint will have on the information we can derive about the filamentary networks around clusters from spectroscopic survey like the WWFCS.

\subsection{Recovery of cluster galaxies after \texttt{Configure}}
\label{sec:memberrecoveryrate}
Physical constraints from the fiber positioner imply that we will never have a $100\%$ completeness of cluster galaxies. Some galaxies won't be targeted as they will appear too close together as well as the constraint of fiber overlap. Therefore, the first test to quantify the success of the WEAVE-like MOS fibre configuration is to estimate the fraction of simulated cluster galaxies with at least one fibre assigned. A high fraction -- particularly outside the cluster core -- will help us achieve our science goals. The overall average fraction of galaxies covered by the WEAVE pointings with at least one fibre allocated (overall target completeness) is $72.7\% \pm 1.7\%$, where the errors denote the scatter of the values for the 160 simulated clusters. If we restrict our calculation to the regions outside the central pointing, which is dominated by the cluster core, the corresponding fraction (outer target completeness) is $81.7\% \pm 1.3\%$. We argue that this value is more relevant than the overall one when dealing with the characterisation of the filament network since the whole cluster core will behave just as a single node \citep{Kuchner20,Kuchner21}. 

We have checked whether the fraction of galaxies selected by configure depends on galaxy mass. If we divide the galaxy sample at the median mass into two equal subsamples, we find that the fraction of high-mass galaxies that are ``configured'' is $\sim77$\%, while the corresponding fraction of low-mass galaxies is $\sim69$\%. This is due to the fact that the central regions of the cluster, which contain a higher fraction of massive galaxies -- high-mass galaxies cluster more strongly than low-mass ones -- are observed twice. Beyond $\sim 2R_{200}$ the fractions are approximately equal.

We find that, whilst the overall target completeness stays relatively constant as a function of cluster mass, the outer target completeness decreases slightly for higher cluster masses. This not surprising since the more massive clusters will have larger cores and a higher surface density of galaxies at all radii. In any case, the sample size reduction induced by the observational constraints seems moderate at all radii. 
 
Note also that the very high number of WEAVE fibres will allow us to observe thousands of background objects per cluster (Figure~\ref{fig:fig5}), providing a thorough test of the accuracy of our photometric redshifts and the quantification of any the possible biases their inclusion in the target selection may introduce (see Kuchner et al., in prep.).

\subsection{Filament network comparison metrics}

We have found that the completeness rates we find are encouragingly high, suggesting that the sample size statistics will not be very severely impacted by the observational constraints. We now need to check whether this sample reduction introduces any biases or changes in the properties of the recovered filamentary networks.

\subsubsection{Skeleton distance}
\label{sed:dskel}
\begin{figure*}
    \centering
    \includegraphics[width = \textwidth]{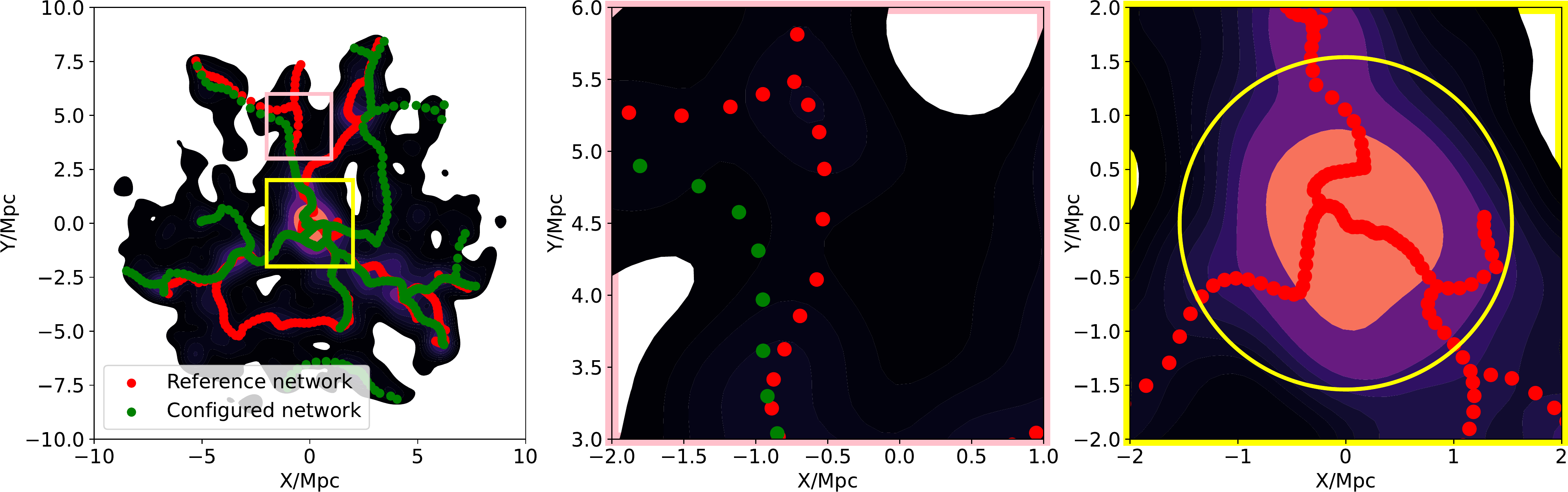}
    \caption{Illustration of the methods used to calculate $D_{\text{skel}}$ and cosmic connectivity. Left panel: the reference filament network (red) and `configured' network (green) are plotted on top of the KDE-smoothed halo density distribution of a simulated cluster, an analogue to WWFCS cluster RX0058. A filament is the amalgamation of many discrete segments, as clearly seen in the middle panel. Middle panel: zoom-in on the region shown by the pink box on the left panel, where the two networks show large differences to demonstrate how $D_{\text{skel}}$ is calculated (see Section \ref{sed:dskel}). Right panel: zoom-in on the cluster core (yellow box in the left panel), only plotting the reference network for illustrative purposes. The circle corresponds to $R_{200}$ and is used to calculate the connectivity as the number of filaments that stem from the main node and cross the $R_{200}$ circle, (see Section \ref{sec:connectivity}). For this cluster, the connectivity has a value of three.}
    \label{fig:fig7}
\end{figure*}

A useful metric designed to quantify the accuracy of filament extraction is the `skeleton distance' $D_{\text{skel}}$ \citep{Laigle18,Malavasi17,Florian}. After running the cosmic web extractor software (\texttt{DisPerSE} in our case), we obtain a series of segments that delineate the cosmic filamentary structure. When comparing two different networks derived in the same region of space, $D_{\text{skel}}$ measures the distance between the start of a segment in the reference network and the nearest one in the other network. The segments that are found are much smaller than the length of a typical filament, allowing us to use the position of the start of a segment as a proxy of a segments position. This is illustrated in Figure~\ref{fig:fig7}. The left panel shows two filamentary networks,  the reference one derived from the full simulated cluster galaxy sample in red, and the `configured' network recovered from the mock-observed galaxy sample in green. The middle panel shows an enlarged version of the pink-boxed region of the left panel, where the differences between the red and green networks are largest, showing the individual segments. The right panel illustrates how we calculate cluster connectivity and is discussed in \ref{sec:connectivity}. The $D_{\text{skel}}$ values are calculated for each segment in the reference network by finding the distance to the nearest segment in the `configured' network. Note that the calculation can also be done in the opposite direction, starting from the segments in the `configured' network instead, and the distribution of $D_{\text{skel}}$ values will not be necessarily the same (see below). In both cases, the distribution of $D_{\text{skel}}$ values quantifies how well both filament networks match each other. 

\begin{figure*}
    \centering
    \includegraphics[width = \textwidth]{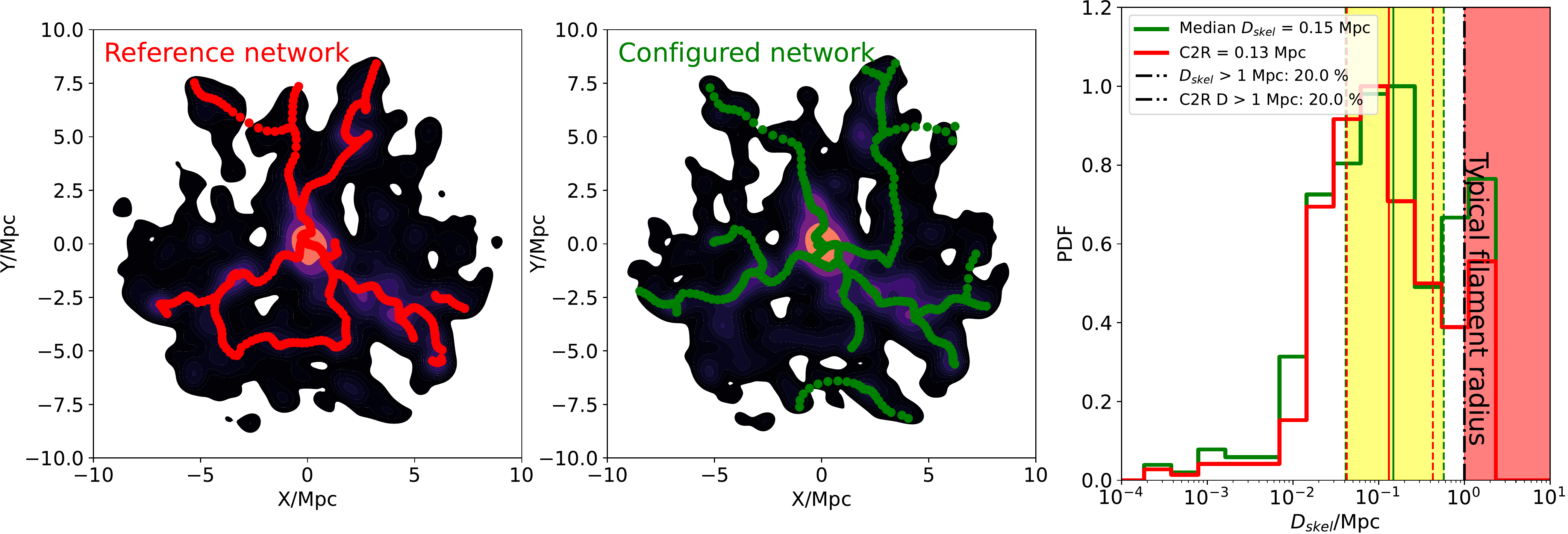}
    \caption{Illustration of the recovery of the filament network around a simulated analogue of the RX0058 cluster, the same cluster as Figure \ref{fig:fig7}. Left panel: KDE-smoothed density distribution of the simulated cluster galaxies with reference filament network in red. Middle panel: as the left panel, but showing the density distribution and filament network (green line) recovered from the `configured' (mock-observed) galaxy sample.  Right panel: $D_{\text{skel}}$ distribution function obtained going from the reference network to the `configured' network (R2C) and vice-versa (C2R), as described in the text. The thick vertical line represents the medians of each distribution, while the shaded yellow region correspond to the 25th and 75th percentiles. The dot-dashed black line represents the typical radius of a filament ($\sim1\,$Mpc). The values of the medians and the percentage of segments with $D_{\text{skel}} > 1\,$Mpc are shown in the legend. We normalized the PDF's to have a common peak value.}
    \label{fig:fig8}
\end{figure*}

Figure~\ref{fig:fig8} shows the reference network (left panel) and the `configured' network (middle panel) for one of the simulated clusters mass-matched to one of the WWFCS target clusters, RX0058. The right panel shows the normalised probability distribution function for $D_{\text{skel}}$, calculated going from the reference network to the `configured' one (R2C, in green) and vice-versa (C2R, in red). The median values are indicated. Both medians are much smaller than the typical radius of filaments ($\sim1\,$Mpc; \citealt{Kuchner20}). A large proportion of $D_{\text{skel}} > $ 1 Mpc would indicate that a filament in this cluster has no counterpart in the corresponding mock-observational cluster. Note that, typically, the median $D_{\text{skel,R2C}}$ is smaller than the median $D_{\text{skel,C2R}}$ because there are generally more segments in the reference network than in the `configured' one, and thus the likelihood of finding a nearer corresponding segment is higher in the R2C direction. If both networks are very similar, both $D_{\text{skel}}$ median values will not only be very small, but also very similar to each other. Therefore, the median values of $D_{\text{skel}}$ and their ratio can be used to quantify the accuracy of filament network reconstruction and also to derive the optimal parameters used by \texttt{DisPerSE} (see Appendix \ref{sec:A3}). 

Another measure of the similarity between the reference and `configured' filament networks is provided by the fraction of $D_{\text{skel}}$ values that are larger than $\sim1\,$Mpc (the typical radius of filaments). The right panel of Figure~\ref{fig:fig8} shows that this fraction is also reassuringly small ($\sim10\%$) in both cases. 

\subsubsection{Cluster connectivity}
\label{sec:connectivity}
Another useful parameter to quantify the accuracy of the filament network derived from the mock observations is the cluster connectivity $C$. We define connectivity as the number of filaments that stem from the main node (cluster core) and terminate beyond $R_{200}$ away from the cluster centre. This definition is slightly different from that of \cite{Laigle18}, where the authors use the cluster virial radius instead of $R_{200}$. The last panel in Figure~\ref{fig:fig7} gives an example of how $C$ is calculated -- there are three filaments stemming from the main node of the network (cluster core) that cross the circle with $R_{200}$ radius, resulting in a cluster connectivity of three. A weak positive correlation between cluster connectivity and cluster mass has been reported in the literature \citep{Florian, Gouin, Darragh-Ford_19, Kraljic_20}, albeit with considerable scatter. Our simulated clusters show a similar correlation.
If the recovered filament network is similar to the reference one, their connectivity should be the same. Therefore, comparing network connectivities will also allow us to assess the accuracy of the recovered filaments.

\subsection{Quantifying the quality of the recovered filament networks}

\begin{figure*}
\centering
    \includegraphics[width = \textwidth]{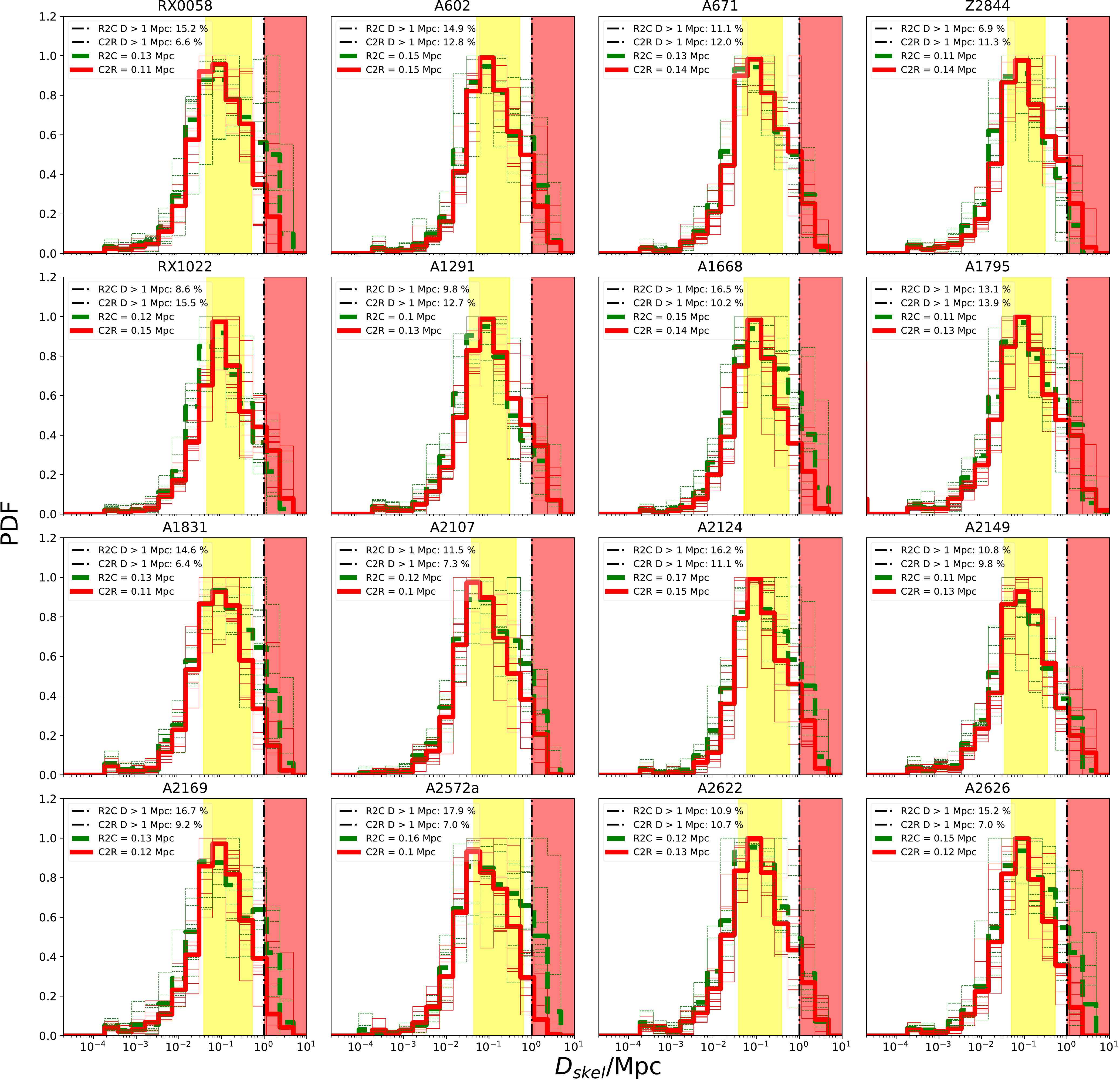}
    \caption{$D_{\text{skel}}$ distributions for all the simulated WWFCS cluster analogues. Each panel shows the individual cluster comparison (thin lines) and the average for the 10 simulated cluster mass-matched to each WWFCS cluster (thick lines). The format of each panel follows that of the right-hand panel of Figure~\ref{fig:fig8}. There is little variation in $D_{\text{skel}}$ over different WWFCS analogue clusters and the $D_{\text{skel}}$ median is always much less than a typical filament radius of 1Mpc.}
    \label{fig:fig9}
\end{figure*}

\begin{figure*}
    \centering
    \includegraphics[width = \textwidth]{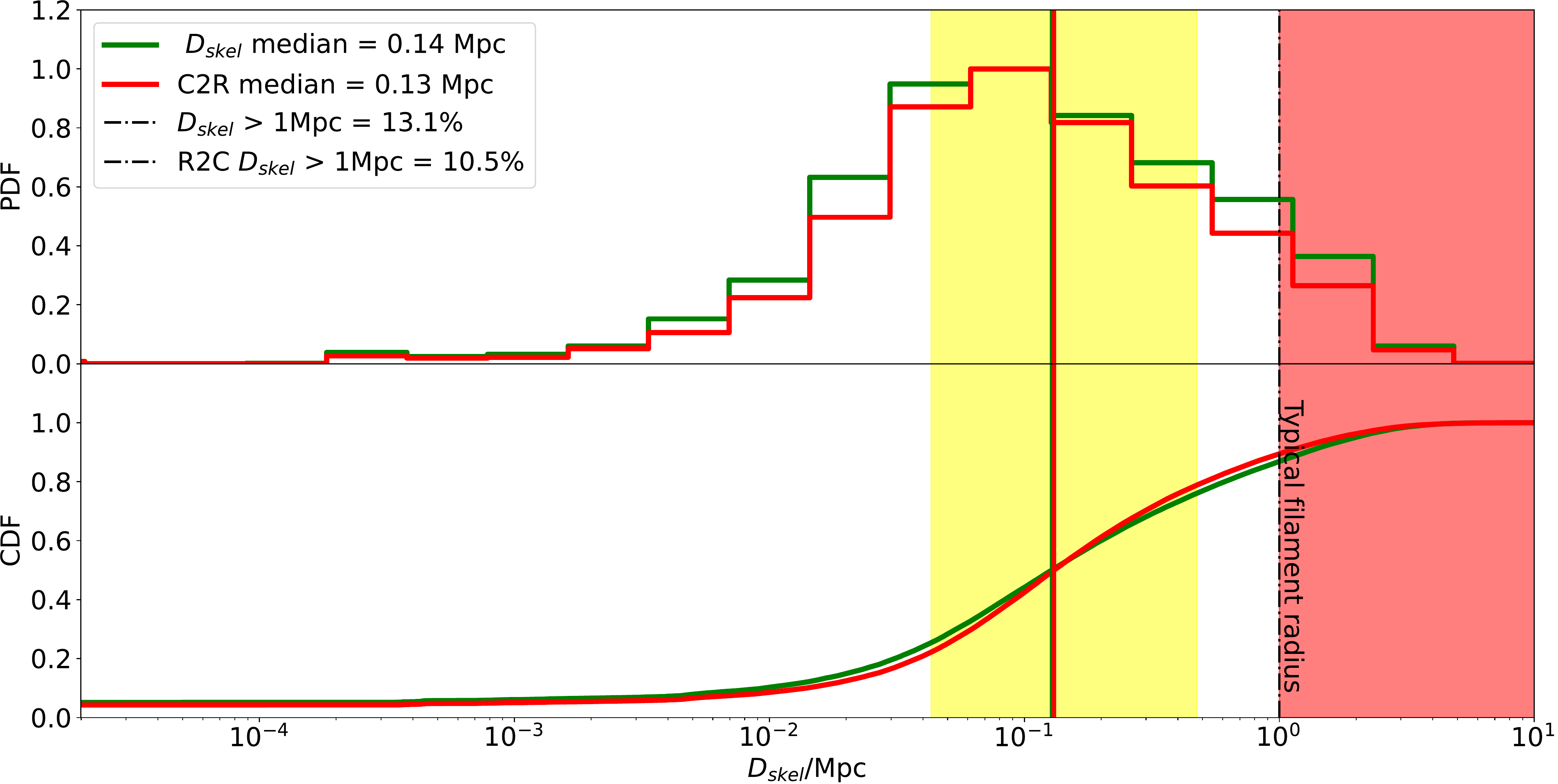}
    \caption{Stacked probability density function (top panel) and cumulative density function (lower panel) of the $D_{\text{skel}}$ distributions corresponding to all the simulated clusters shown in 
    Figure~\ref{fig:fig9}. The positional difference in the networks is minimal -- our mock observations of filaments around galaxy clusters successfully recreate the 'true' simulated filament network.}
    \label{fig:fig10}
\end{figure*}
We are now in a position to use the $D_{\text{skel}}$ and connectivity metrics to assess quantitatively the impact of the WWFCS observational strategy and constraints on the recovery of the filament networks surrounding galaxy clusters. 

As described above, Figure~\ref{fig:fig8} illustrates the filament network comparison process for a single simulated cluster, mass-matched to RX0058. Visually, there is remarkable similarity in the reference and `configured' filament networks. For this particular cluster, $71.8\%$ of the simulated cluster galaxies have at least one MOS fibre allocated, which is close to the average for the whole sample. The median values of $D_{\text{skel}}$ are $0.12\,$Mpc going in the R2C direction and $0.16\,$Mpc going in the C2R direction (cf. Section~\ref{sed:dskel}). These values are much smaller than $\sim1\,$Mpc, the typical radius of filaments. Moreover, only $8\%$ and $10\%$ of the filamentary segments lie at a distances greater than $\sim1\,$Mpc.

The cluster connectivity of the reference network is 3, while one of the filaments in the `configured' network bifurcates inside $R_{200}$, increasing the connectivity to 4. Changes in connectivity of $\pm1$ are not uncommon, indicating that the recovery is not perfect. However, larger changes in connectivity are rare (see below).  

These results, if replicated for the whole cluster sample, are very encouraging, suggesting that the data provided by the WWFCS will allow the reliable recovery of the filamentary structures around clusters since the impact of the observational constraints will be moderate. 

Figure~\ref{fig:fig9} confirms that the $D_{\text{skel}}$ results shown for the RX0058 analogue are indeed typical of the whole sample. We can therefore stack the $D_{\text{skel}}$ distributions for the 160 simulated clusters (Figure~\ref{fig:fig10}) and derive representative average quantities for the whole sample. On average, the median values of $D_{\text{skel,R2C}}$ and $D_{\text{skel,C2R}}$ are $0.13\pm0.02\,$Mpc. The values are not only reassuringly small, but also almost exactly the same when going in both directions, strongly suggesting the compared filament networks are very similar. Furthermore,  typically only $11$--$13\%$ of the corresponding filamentary segments are more that  $1\,$Mpc away from each other.

\begin{figure}
    \centering
    \includegraphics[width = 0.5\textwidth]{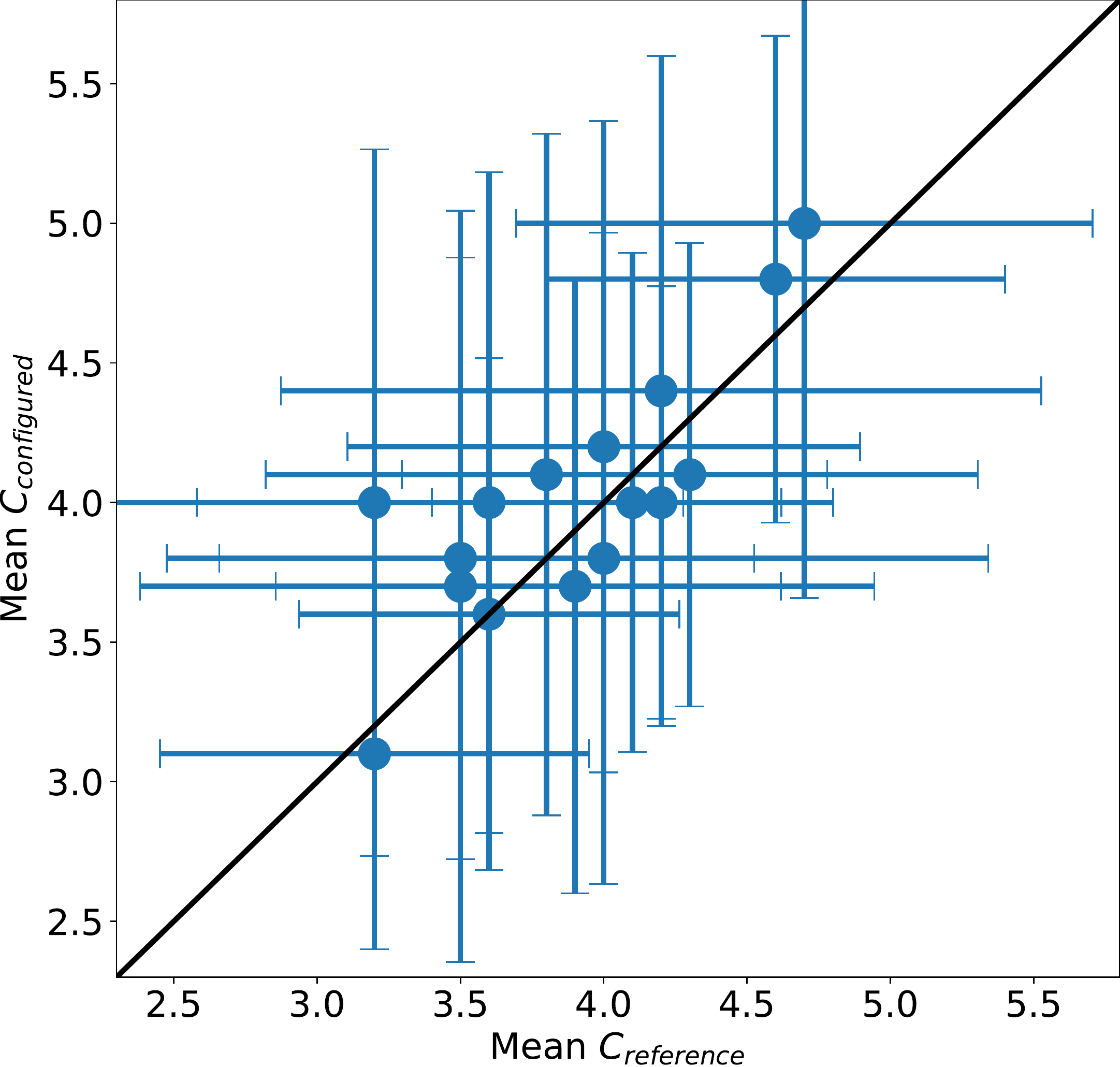}
    \caption{Comparison between the cluster connectivity $C$ of the reference and `configured' filament networks. The 1-to-1 line is shown in black. Each point corresponds to the mean connectivity of the 10 mass-matched analogues of each WWFCS cluster, and the error bars show the $\pm1\sigma$ scatter.} 
    \label{fig:fig11}
\end{figure}

Figure~\ref{fig:fig11} shows a comparison between the connectivity of the reference filament networks $C_\text{reference}$  and that of the `configured' ones $C_\text{configured}$. The mean values corresponding to each WWFCS match very well, with no significant bias, and the small scatter indicates changes in $C$ are generally kept within the $\pm1$ range. 

In summary, the quantitative tests we have carried out for the whole cluster sample confirm our initial assessment that the impact of the observational constraints imposed by the WWFCS on the recovery of the filament networks around galaxy clusters will be very moderate.

\section{Summary and Conclusions}

The outskirts and infall regions of galaxy clusters act as the points of contact linking the large-scale structure of the Universe to the highly dense virialized cores of the clusters themselves, containing some of the key environments affecting galaxy evolution. 
Since next generation spectroscopic surveys such as the Weave Wide Field Cluster Survey (WWFCS\footnote{Although in this paper we have focused on the WWFCS, our methodology could easily be adapted and applied to other wide-field spectroscopic surveys.}) will explore and map in detail these complex regions, in this paper we forecast how successful such surveys will be at identifying the filaments that link together the `nodes' in the large-scale structure -- clusters and groups -- and channel galaxies into them.

We aim at quantifying the impact the observational limitations will have on our ability to detect the filamentary structures that feed the clusters in the WWFCS. To achieve that aim we have used a large sample of simulated massive galaxy clusters from \texttt{TheThreeHundred} project \citep{Cui2018} and created a set of simulated cluster galaxy samples closely matching the selection and observational constraints imposed by the WWFCS (Kuchner et al., in prep.). For each one of the 16 WWFCS target clusters we have extracted 10 mass-matched analogue clusters from the simulations and built mock-observed galaxy samples reaching beyond $\sim5R_{200}$, where cosmic filaments trace and connect galaxy clusters to the cosmic web. We summarise our main results below. 

\begin{enumerate}
  \item We have then followed closely the strategy, selection, and observational constraints of the WWFCS. Applying the same MOS fibre configuration tool that the WEAVE spectrograph will use, we find that, on average, we are able to allocate fibres to $72.7\% \pm 1.7\%$ of all the target galaxies. More importantly, outside the cluster core -- in the outer regions that are crucial for filament identification -- the success rate increases to $81.7\% \pm 1.3\%$. The number of cluster galaxies that are targeted ranges from 1284 -- 4062. The high completeness that the WEAVE observations will allow, together with the large field coverage, are key to the success of the survey. \\
  \item In each of the simulated cluster regions we have used the filament finder \texttt{DisPerse} \citep{Sousbie11} to trace the cosmic-web filament skeleton before and after the observational constraints (including MOS fibre positioning) are imposed on the galaxy samples. We then compared quantitatively the resulting filament networks and find that we are able to recover the original network remarkably well. Specifically, we find that the median distance between corresponding filament segments $D_{\text{skel}}$ in the reference and recovered networks  is only $0.13 \pm 0.02\,$Mpc on average, an much smaller than the typical filament radius of $\sim1\,$Mpc. Furthermore, only $\sim11-13\%$ of all recovered filament segments lie at a distance larger than $1\,$Mpc away from their corresponding reference segment. \\
  \item As a further test on the integrity of the recovered filament networks we computed their \textit{connectivity}, the number of filaments that stem from the cluster core and terminate beyond $R_{200}$ away from the cluster centre. We find that the connectivities of the reference and recovered networks match very well, without any significant bias, indicating that their global properties are also recovered well.  \\
 \end{enumerate}

These findings make us confident that the WWFCS will be able to reliably trace cosmic-web filaments in the vicinity of massive galaxy clusters. The next step, when we start receiving data from WEAVE, will be to identify the galaxies that belong to these filaments, and compare their properties (e.g., mass, metallicity, star formation, stellar populations) to those of galaxies inhabiting other environments such as groups, the clusters cores, and the general field. With the combination of a statistical sample of clusters together with high target sampling rate, the WWFCS will provide a detailed look at the influence of all environments in the cluster infall regions on galaxy evolution.

\section*{Acknowledgements}

Funding for the WEAVE facility has been provided by UKRI STFC, the University of Oxford, NOVA, NWO, Instituto de Astrofísica de Canarias (IAC), the Isaac Newton Group partners (STFC, NWO, and Spain, led by the IAC), INAF, CNRS-INSU, the Observatoire de Paris, Région Île-de-France, CONCYT through INAOE, Konkoly Observatory of the Hungarian Academy of Sciences, Max-Planck-Institut für Astronomie (MPIA Heidelberg), Lund University, the Leibniz Institute for Astrophysics Potsdam (AIP), the Swedish Research Council, the European Commission, and the University of Pennsylvania.  The WEAVE Survey Consortium consists of the ING, its three partners, represented by UKRI STFC, NWO, and the IAC, NOVA, INAF, GEPI, INAOE, and individual WEAVE Participants. Please see the relevant footnotes for the WEAVE website\footnote{\url{https://ingconfluence.ing.iac.es/confluence//display/WEAV/The+WEAVE+Project}} and for the full list of granting agencies and grants supporting WEAVE\footnote{\url{https://ingconfluence.ing.iac.es/confluence/display/WEAV/WEAVE+Acknowledgements}}.

For the purpose of open access, the authors have applied a creative commons attribution (CC BY) to any journal-accepted manuscript.

This work has been made possible by \texttt{TheThreeHundred} collaboration, which benefits from financial support of the European Union’s Horizon 2020 Research and Innovation programme under the Marie Sk\l{}odowskaw-Curie grant agreement number 734374, i.e. the LACEGAL project. \texttt{TheThreeHundred} simulations used in this paper have been performed in the MareNostrum Supercomputer at the Barcelona Supercomputing Center, thanks to CPU time granted by the Red Espa\~{n}ola de Supercomputaci\'{o}n. 

DC acknowledges support from STFC through a studentship. He thanks Kenneth Duncan, Sarah Hughes and Gavin Dalton for helping setup and manipulating the software \texttt{Configure}. He also thanks Benedetta Vulcani and Daniela Bettoni for useful feedback on the contents of the paper. He thanks Tomas Hough for providing the SAG galaxy property files. UK acknowledges support from the Science and Technology Facilities Council through grant number RA27PN. WC is supported by the STFC AGP Grant ST/V000594/1 and the Atracci\'{o}n de Talento Contract no. 2020-T1/TIC-19882 granted by the Comunidad de Madrid in Spain. He further acknowledges the science research grants from the China Manned Space Project with NO. CMS-CSST-2021-A01 and CMS-CSST-2021-B01. LPdA acknowledges financial support from grant PGC2018-094671-B-I00 funded by MCIN/AEI/10.13039/501100011033 and by the European Union NextGenerationEU/PRTR.
The authors contributed to this paper in the following ways: DC, UK, AAS, MEG and FRP formed the core team. DC analysed the data, produced the plots and wrote the paper along with UK, AAS and MEG. 

\section*{Data Availability}
Data available on request to \texttt{TheThreeHundred} collaboration: https://www.the300-project.org.

\bibliographystyle{mnras}

\bibliography{bibliography} 

\appendix
\section{Process of optimizing the WWFCS field positions}
\label{sec:A1}
As mentioned in section ~\ref{sec:fields}, the WWFCS performs observations by arranging 2 degree fields into a mosaic pattern, covering the cluster core, infall region and outskirts (Figure ~\ref{fig:fig1}). To optimize the observational strategy, we aim to design the field positions in a way that maximizes the cluster coverage. Firstly, we place two fields at the core of each cluster, the region of the highest number density. This is so we can maximise the number of targeted cluster members over the total field of view. To optimize the tiling for each cluster, we adopt the following regime, such that if
\begin{equation}
    \frac{\text{Area within } 5R_{200} \text{ for N - 1 fields}}{\text{Area within } 5R_{200} \text{ for N fields}} > 97\%,
\end{equation}
\begin{figure}
    \centering
    \includegraphics[width = 0.5\textwidth]{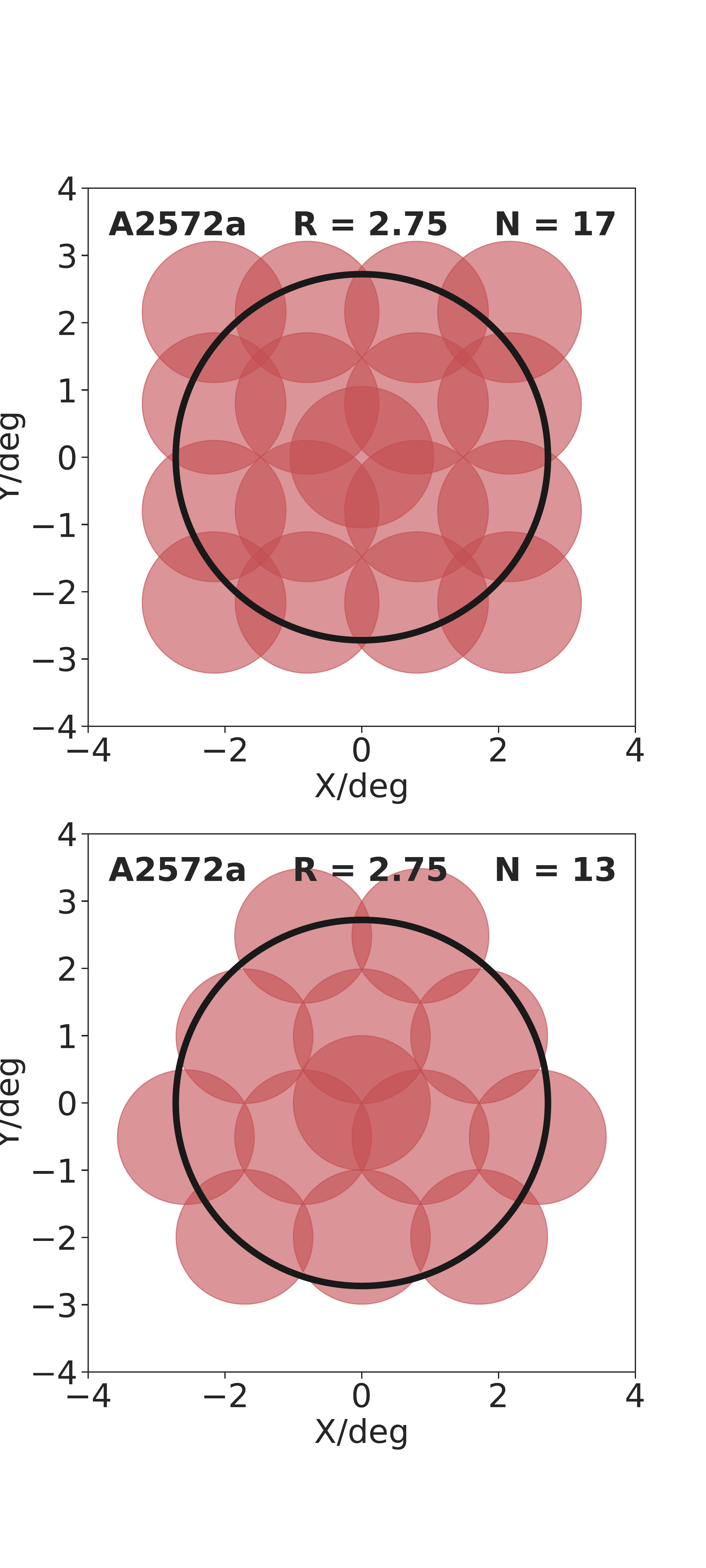}
    \caption{An example field layout of cluster A2572a before this work (top) and after this work (bottom). The red circles are individual 2-degree diameter WEAVE fields whilst the black outer circle represents the angular diameter corresponding to $5R_{200}$ of this cluster. The numbers displayed are the total number of fields required to cover this cluster $N$, the cluster redshift and $R_{200}$ taken from \citep{Omegawings}.}
    \label{fig:figA1}
\end{figure}
\begin{figure*}
\centering
    \includegraphics[width = 0.85\textwidth]{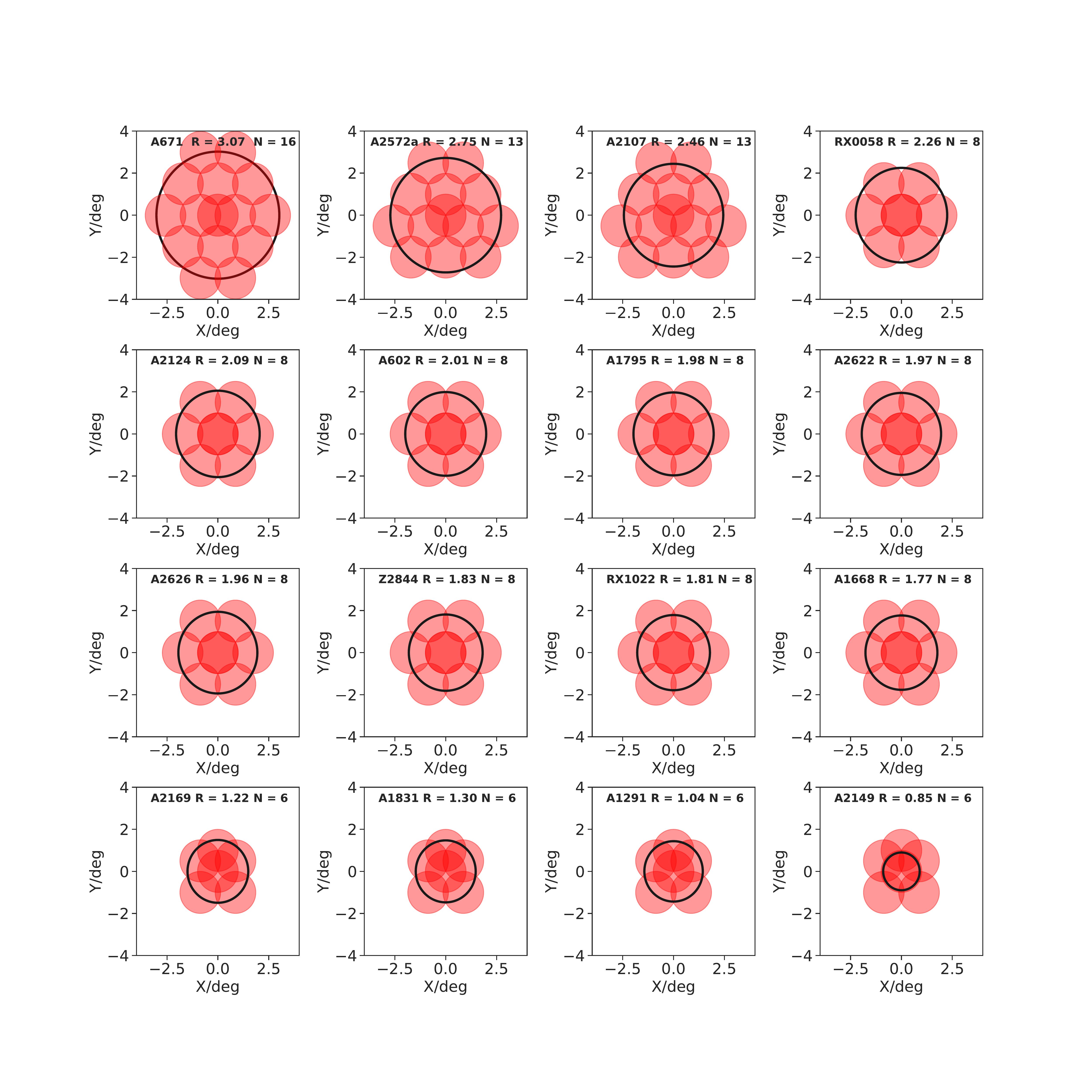}
    \caption{WEAVE field pattern for each WWFCS cluster. The caption in each panel states the name of the cluster, $5R_{200}$ in degrees and the number of fields $N$ used for each observation.}\label{fig:a}
    \label{fig:figA2}
\end{figure*}
then we can remove one field, (we use 97\% to ensure that we are still covering a significant area within $5R_{200}$). We iterate through this process by removing fields in the outer region of the cluster, manually inspecting each time one is removed, until the 97\% threshold is exceeded.
Starting from a 'naive' geometric tiling pattern (illustrated in Figure \ref{fig:figA1}, top), the total number of WEAVE fields required to cover the 16 clusters was 155, adding up to $147\,250$ fibre hours. Using the new optimised tiling method (illustrated in Figure \ref{fig:figA1}, bottom) the total number of fiber hours is reduced to $130\,390$. For the example shown in Figure \ref{fig:figA1}, even though we have removed four fields, we are still covering out to $5R_{200}$. Of the clusters, 12 out of 16  have full coverage out to $5R_{200}$, whilst overall we have lost a total of $0.06\%$ area within $5R_{200}$. 
\section{Galaxy cluster scaling}
\label{sec:A2}
This sections details our method for scaling down the mass of the clusters from  \texttt{TheThreeHundred} to match the WWFCS selected clusters, as mentioned in Section ~\ref{sec:mass_match}.

\begin{figure*}
    \centering
    \includegraphics[width = 0.85\textwidth]{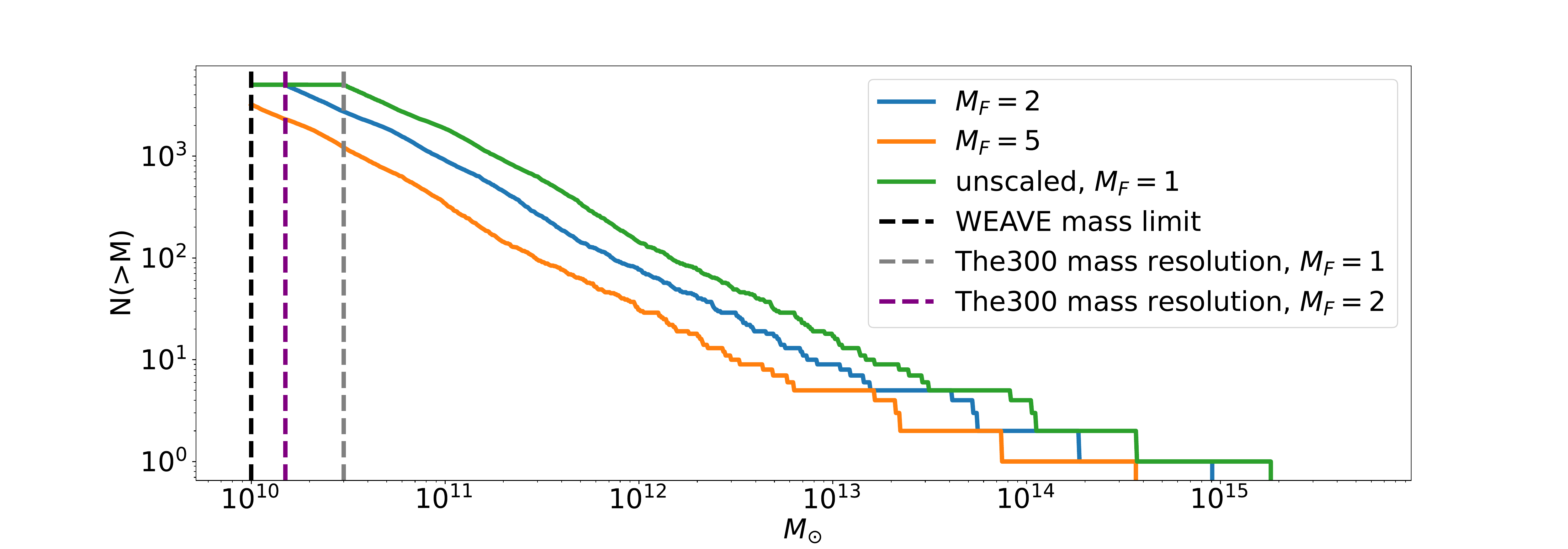}
    \caption{Cumulative sum of individual halo masses of simulated galaxies for one galaxy cluster in \texttt{TheThreeHundred} that has been modified using the three $M_{F}$. All haloes follow the criterion set for selecting 'high--quality data' from the simulations, as described in Section ~\ref{sec:data_300}. Different colours represent different scale factors that have been applied to the simulated catalogues. The vertical dotted lines represent different mass resolutions.}
    \label{fig:figB1}
\end{figure*}

Firstly, we arbitrarily chose three mass-scaling factors: $M_{F} = 1, 2$ and 5, which the simulated cluster mass is divided by. We chose the most-massive halo in the corresponding cluster catalogue to act as a proxy for the cluster. Also, we increase our cluster catalogue sample size by a factor of 3 by including each 2D-plane, (xy, xz, yz). The resulting mass distributions of the clusters that have been scaled down by $M_{F}$ are in Figure~\ref{fig:fig3}, where we have demonstrated that by choosing these mass factors, we have covered the entire WEAVE mass range. 

For each scaling factor $M_F$, we divide the mass range into 20 mass bins.To create a statistically significant sample, we draw from these bins with the aim of identifying 10 mass-matched simulated analogue clusters for each of the 16 WEAVE clusters. Our total sample of analogue clusters is thus 160. The presence of companion clusters within $5 < R < 15$ Mpc of the WWFCS clusters (the radius of the simulation volume) is unknown, therefore we do not exclude analogue clusters with secondary clusters within this distance. Within the sample of 160 analogue clusters there are six such configurations. 

In scaling the mass of the clusters from the simulations to match WEAVE, we have to also individually scale the mass of all the associated halos for each cluster. Figure~\ref{fig:figB1} displays the cumulative number of simulated halos that lie above a mass interval for one cluster. We see that for higher mass scaling factors ($M_{F}$) we are shifting the masses of all the haloes associated with the cluster to lower values.

We are limited in our recovery of haloes by two thresholds: The 'scaled' simulation mass resolution and the observational mass resolution. In the case where we don't scale the simulations, ($M_{F} = 1$), the mass resolution limit is that of the dark matter particles in the simulations \citep{Kuchner21}, given by the dotted grey line in Figure \ref{fig:fig5}. For $M_F > 1$, we reduce the simulation mass resolution by dividing it by the mass scaling factor $M_{F}$. Whilst we can change the simulated mass resolution threshold, there is a hard limit on the observational mass limit. For $F > 3$, the mass resolution stays at the observational limit $M_{\text{obs}} = 10 \times 10^{10} M_{\odot}$, (corresponding to the r band limit used for WEAVE: $r_{\text{total}} < 19.75$ which is equivalent to a stellar mass of ~ $10^{9} M_{\odot}$, (Kuchner et al., in prep)). At $M_{F} > 3$, we artificially lose halos that are massive enough to be simulated, but are too small to meet the WWFCS observational criteria. However, as shown by the top panel in Figure~\ref{fig:fig3}, we are still obtaining thousands of cluster galaxies per cluster. This reduction in the cluster member population mimics the expected relation of lower mass cluster's hosting fewer subhaloes (see for example \citealt{Poggianti2010}). 

\section{DisPerSE input parameters}
\label{sec:A3}
\begin{figure*}
    \centering
    \includegraphics[width = \textwidth]{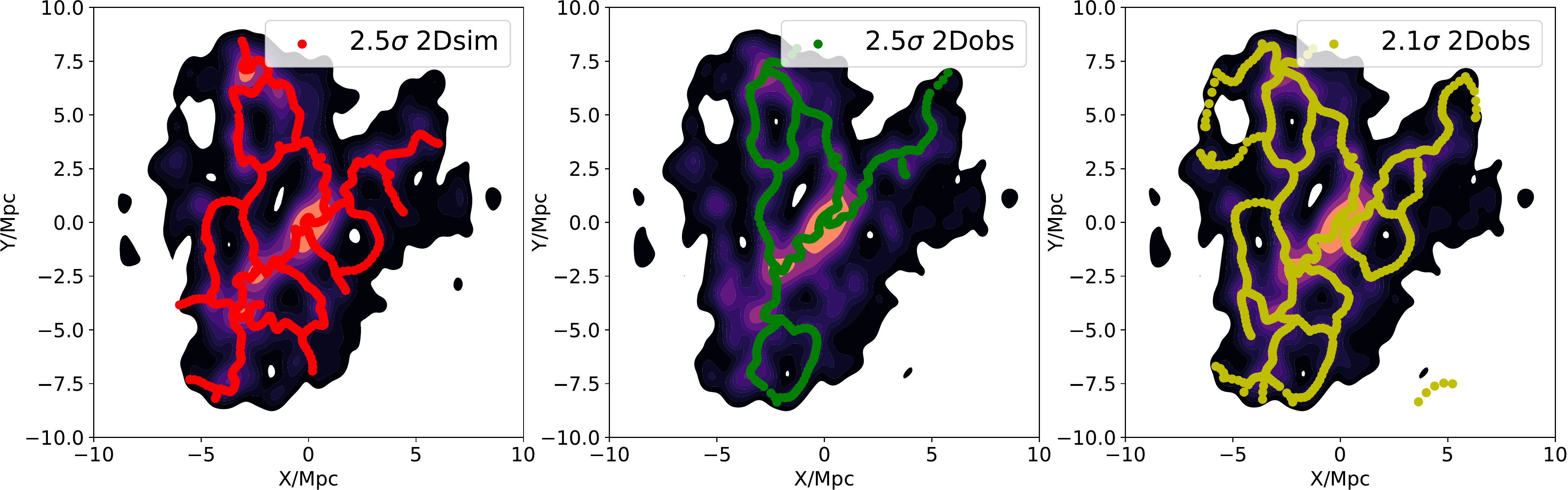}
    \caption{This figure illustrates how  the change in the density field when `configuring' a cluster makes a change in the persistence parameter necessary when extracting the filament network. Left panel: KDE-smoothed density field of a simulated cluster analogue to WWFCS target RX0058, with the filaments identified by the topological structures extractor DisPerSE traced on top (using our choice of persistence for the `pre-configured' model, $2.5\sigma$. Middle panel: the `mock-observational' cluster with filaments extracted using the same persistence, $2.5\sigma$. Right panel: the same but with our choice of persistence for the `configured' model, $2.1\sigma$. Lowering the persistence yields a more accurate reconstruction of the cosmic web around galaxy clusters.}
    \label{fig:figC1}
\end{figure*}
As mentioned in Section~\ref{sec:config}, to extract the filament networks with DisPerSE we need to set a persistence threshold. We introduce a metric $\Psi$ with 3 key parameters:\\
1) $D_{\text{skel}}$ median (the positional difference in the reference network spine and the configured network spine),\\
2) $D_{\text{skel}}$ ratio: the ratio of the two methods of calculating $D_{\text{skel}}$,\\
3) Cluster connectivity of 'configured' network ($C$, number of filaments that stem from the main node and terminate outside $R_{200}$) comparison to reference network.\\

All of the above parameters are normalized against their maximum output given a persistence and are equally weighted. We compute this metric in a suitable range of persistences $2<\sigma<3$, in 0.1$\sigma$ intervals, and the persistence that minimizes $\Psi$ returns the most accurate reconstruction of the filamentary network. A $\Psi$ of three corresponds to the worst possible reconstruction of the network, whilst $\Psi$ values close to zero represent the most accurate filamentary mapping. We have automated a process in determining the best persistence given a reference skeleton. Our scientific rationale requires a high completeness, therefore, we explore low persistence values that not only map out the most robust structure, but also filaments that connect nodes with smaller persistence ratios. After analysing $\Psi$ for different networks for different clusters, we selected a persistence of $\sigma = 2.5$ for the reference network and $\sigma = 2.1$ for the configured network. In the process of `configuring' a cluster, we are effectively altering the underlying density field and therefore, it is necessary to change the input persistence, as demonstrated in Figure \ref{fig:figC1}. Although varying the persistence cluster-by-cluster can change $D_{\text{skel}}$, the median and PDFs of $D_{\text{skel}}$ do not vary significantly with changes in persistence when averaged over all 160 clusters.

\bsp	
\label{lastpage}
\end{document}